\pdfoutput=1

\documentclass[twocolumn,showpacs]{revtex4}
\usepackage{graphicx,amssymb,amsmath}
\usepackage[usenames,dvipsnames]{color}    
\begin{document}

\title{
  Random matrices and quantum chaos in weakly-disordered
  graphene nanoflakes
}

\author{Adam Rycerz}
\affiliation{Marian Smoluchowski Institute of Physics, 
Jagiellonian University, Reymonta 4, PL--30059 Krak\'{o}w, Poland}

\begin{abstract}
Statistical distribution of energy levels for Dirac fermions confined in a quantum dot is studied numerically on the examples of triangular and hexagonal graphene flakes with random electrostatic potential landscape. When increasing the disorder strength, level distribution evolves from Poissonian to Wigner, indicating the transition to quantum chaos. The unitary ensemble (with the twofold valley degeneracy) is observed for triangular flakes with zigzag or Klein edges and potential varying smoothly on the scale of atomic separation. For small number of edge defects, the unitary-to-orthogonal symmetry transition is found at zero magnetic field. For remaining systems, the orthogonal ensemble appears. These findings are rationalized by means of additive random-matrix models for the cases of weak and strong intervalley scattering of charge carriers in graphene. The influence of weak magnetic fields, as well as the strong-disorder-induced wavefunction localization, on the level distribution is also briefly discussed.
\end{abstract}

\date{\today}
\maketitle

\section{Introduction}
The advent of graphene \cite{Nov05}, an experimental model for two-dimensional massless Dirac fermions \cite{Div84}, has provided the unique opportunity the test the theoretical predictions for such exotic particles from one perspective \cite{Bee08}, and to reexamine the classic effects of nanoscopic physics \cite{Naz09} from the other perspective. Theoretical predictions already verified experimentally include Klein tunneling \cite{Kat06a} pseudodiffusive shot noise \cite{Two06,Dan08}, the universal quantum values of the conductivity \cite{Two06,Dan08,Kat06b} and the visible light opacity \cite{Nai08}. On the other hand, nanostructures in graphene show weak localization \cite{Mac06}, universal conductance fluctuations \cite{Ryc07}, Aharonov-Bohm \cite{Rec07,Rus08,Wur10} and Josephson \cite{Tit06} effects, just to mention a~few.

These two perspectives unify in the issue of quantum chaos in Dirac billiards \cite{Ber87}, modelled experimentally within the graphene quantum-dot devices \cite{Pon08}. Early theoretical considerations \cite{Ber87} suggested that in Dirac billiards {\em time-reversal symmetry} (TRS) may be broken even in the absence of magnetic fields, leading to the {\em unitary} symmetry class. Existing Coulomb-blockade experiments \cite{Pon08} provide strong indications for quantum chaos in graphene, but without giving a clear identification of the symmetry class. Several computational experiments were performed \cite{Wur09,Der08,Lib09} showing, that level-spacing distributions for irregular or disordered graphene nanoflakes exhibit {\em orthogonal} symmetry class, as scattering the carriers between $K$ and $K'$ valleys restores TRS \cite{Wur09}. In turn, when searching for the unitary symmetry, one should focus rather on {\em open} than {\em closed} nanosystems in graphene. 

\begin{figure}[!b]
\centerline{\includegraphics[width=0.9\linewidth]{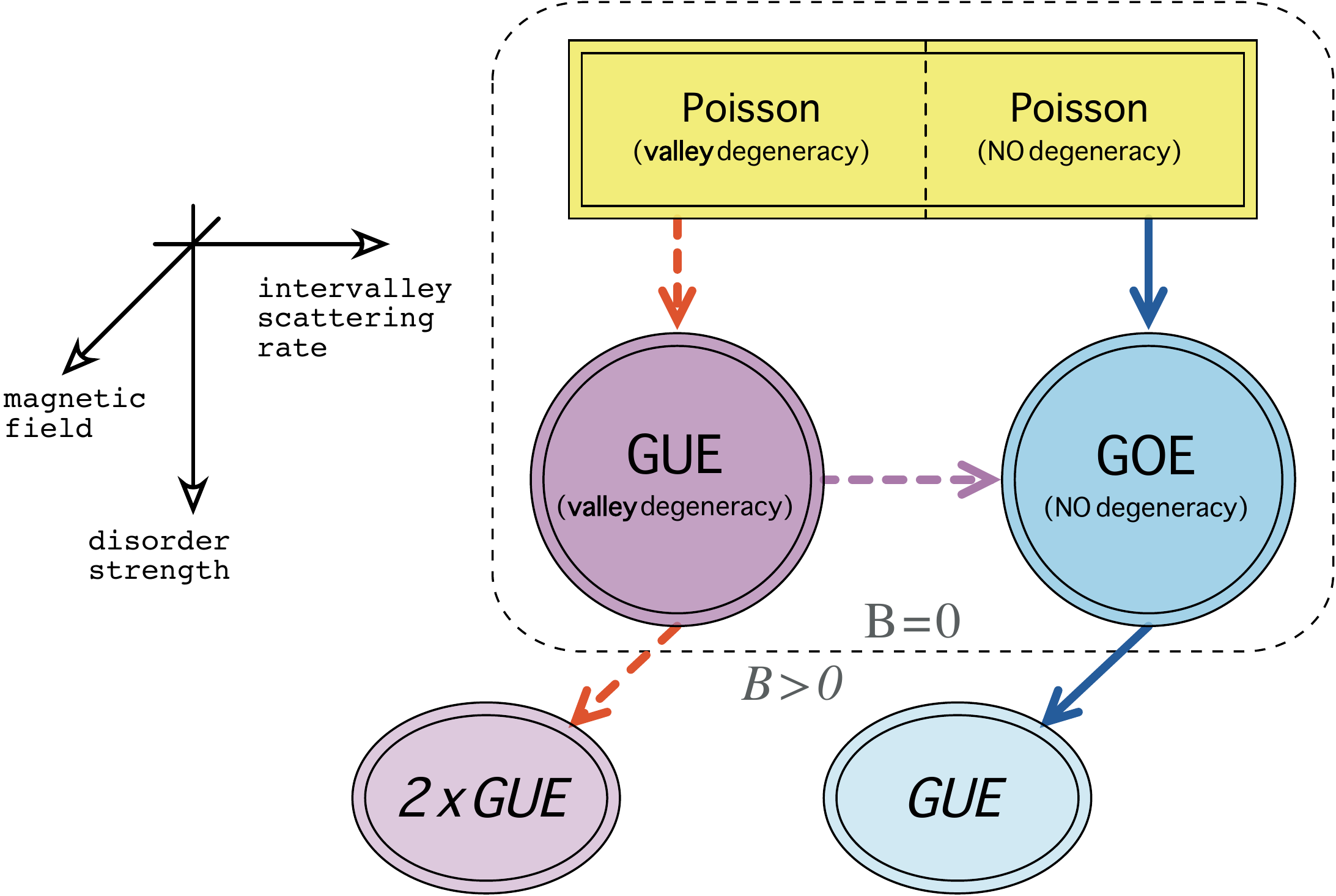}}
\caption{\label{gradyn}
Transitions between symmetry classes and random matrix ensembles relevant for {\em closed} nanosystems in graphene characterized by the disorder strength, the intervalley scattering rate, and (optionally) placed in the weak magnetic field $B$. Solid arrows in the right part indicate transitions already reported in the literature; dashed arrows in the central part indicate remaining transitions. 
}
\end{figure}

In this paper we follow the line of approach established with Refs.\ \cite{Wur09,Der08,Lib09}, but focus on highly-symmetric (triangular and hexagonal) graphene nanoflakes, in which transition to quantum chaos is driven by weak potential disorder \cite{Lew08,Car08,Wur11b,Ryc11} attributed to the influence of substrate impurities (or ions). Our numerical results show, that albeit the orthogonal symmetry appears generically in closed graphene nanosystems, in a~peculiar case of {\em triangular} nanoflakes with zigzag (or Klein \cite{Kle94}) edges and smooth impurity potential the unitary symmetry class is the relevant one. Also, in such a case the (approximate) twofold valley degeneracy is observed, as the scattering of carriers between the valleys is negligibly weak. When turning on the magnetic field, such a system transforms into a~pair of two independent chaotic systems (one at each valley) each of which showing the unitary symmetry. On the other hand, edge defects at zero field increase the intervalley scattering, such that the twofold degeneracy is lifted up and TRS is restored (leading to the orthogonal symmetry class). These findings complement the diagram of possible transitions between symmetry classes of graphene nanoflakes (see Fig.\ \ref{gradyn}).

The structure of the paper is as follows. In Section \ref{didigra}, we discuss the relevant symmetries of the Hamiltonian for Dirac fermions in graphene and present the tight-binding model for a potential disorder. This model is utilized in Section \ref{leretri} to demonstrate the level repulsion appearing in graphene nanoflakes when increasing the disorder strength. Next, in Section \ref{ramaspe}, we overview the basic random-matrix models for dynamical systems and apply them to describe the transition to quantum chaos in graphene nanoflakes of different shapes and boundaries. In Section \ref{tragogu}, we investigate the two distinct transitions between the orthogonal and the unitary symmetry appearing in triangular nanoflakes at zero or finite magnetic fields, and compare the spectral statistics for each case with those obtained from relevant random-matrix models. The influence of disorder-induced localization on spectral statistics is discussed in Section \ref{wlocal}. The conclusions are given in Section \ref{conclu}.

\section{Dirac fermions in weakly disordered graphene\label{didigra}}
The microscopic model of disorder in graphene nanoflakes, representing the random electrostatic potential landscape \cite{Ryc07,Lew08,Car08,Wur11b,Ryc11}, is presented in this Section. Depending on the disorder correlation length $\xi$, the model may represent the potential {\em abruptly} ($\xi\ll{a}$) or {\em smoothly} ($\xi\gg{a}$)  varying on the length-scale of the lattice spacing in graphene $a=0.246\,$nm. But first, let us briefly recall (after Ref.\ \cite{Wur09}) the discussion of possible symmetry classes of such nanosystems.

\subsection{Symmetries of the Hamiltonian}
The effective Hamiltonian for low-energy excitations of electrons in graphene in low magnetic fields \cite{zeeman} has a~form of the Dirac Hamiltonian
\begin{multline}
\label{hameff}
  {\cal H}_{\rm Dirac} = \\
  v_F(p_x\!+\!eA_x)\,\sigma_x\otimes\tau_z + v_F(p_y\!+\!eA_y)\,\sigma_y\otimes\tau_0 \\
  + M(x,y)\,\sigma_z\otimes\tau_0 + U(x,y)\,\sigma_0\otimes\tau_0,
\end{multline}
where $v_F\simeq{10}^6\,$m/s is the energy-independent Fermi velocity, $p_i=-i\hbar\partial_i$ (with $i=1,2$) are in-plane momentum operator components, $\sigma_j$ and $\tau_j$ ($j=1,2,3$) are the Pauli matrices acting on sublattice and valley degrees of freedom (respectively), and $\sigma_0$ ($\tau_0$) denotes the unit matrix. The electron charge is $-e$, the vector potential ${\bf A}=(A_x,A_y)$ defines perpendicular magnetic field via $B_z=\hat{\bf e}_z\cdot\mbox{rot}\,{\bf A}=\partial_xA_y-\partial_yA_x$, whereas $M(x,y)$ and $U(x,y)$ are the mass term and the electrostatic potential energy (respectively). The Hamiltonian (\ref{hameff}) acts on spinors $\psi\equiv[\psi_A,\psi_B,\psi_A',\psi_B']^T$, where $A/B$ is the the sublattice index and the primed and unprimed entries correspond to the two valleys. Two-component wavefunction for a~charge carrier in the position representation $\Psi(x,y)=[\Psi_A,\Psi_B]^T$ is determined by the solution of the Dirac equation ${\cal H}_{\rm Dirac}\psi=E\psi$ via
\begin{equation}
\label{wavefun}
  \Psi({\bf r})=
  \left(\begin{array}{c}\psi_A \\ \psi_B\end{array}\right)e^{i{\bf K}\cdot{\bf r}}+
  \left(\begin{array}{c}\psi_A' \\ \psi_B'\end{array}\right)e^{i{\bf K'}\cdot{\bf r}},
\end{equation}
where ${\bf r}\equiv(x,y)$ and ${\bf K}$ (${\bf K'}$) stands for the position of $K$ ($K'$) valley in the momentum space. (We choose ${\bf K}=-{\bf K'}=\frac{2\pi}{3a}\hat{\bf e}_x$ for the remaining parts of the paper.) Eq.\ (\ref{wavefun}) allows one to discuss the position dependence of each spinor component of $\psi$, slowly varying on the scale of atomic separation $a$ (apart from the full wavefunction $\Psi({\bf r})$ varies abruptly on the scale of $a$).

Symmetries of the Hamiltonian (\ref{hameff}) are defined by the following antiunitary operations: standard time reversal ${\cal T}$, and two ``special time reversals''
\begin{gather}
  {\cal T}= (\sigma_0\otimes\tau_x){\cal C},   \\
  {\cal T}_{\rm sl}= -i(\sigma_y\otimes\tau_0){\cal C}, \ \ \ 
  {\cal T}_v= -i(\sigma_0\otimes\tau_y){\cal C}, \label{tsltv}
\end{gather}
where ${\cal C}$ denotes complex conjugation. The mass term $M(x,y)\sigma_z\otimes\tau_0$ breaks the {\em symplectic symmetry} associated with ${\cal T}_{\rm sl}$, leading to the two distinct possible scenarios: 
\begin{enumerate}
\item[(i)] 
In the case of {\em weak intervalley scattering}, ${\cal T}_v$ commutes with ${\cal H}_{\rm Dirac}$, so the system consists of two independent subsystems (one for each valley). Each subsystem lacks TRS (even at zero magnetic field), as ${\cal T}$ commutes only with full ${\cal H}_{\rm Dirac}$. Because the Kramer's degeneracy (${\cal T}_v^2=-I$) \cite{Haa10}, the Hamiltonian of a~chaotic system consists of two degenerate blocks (one per each valley), each of which may be modelled by a random matrix belonging to the {\em Gaussian Unitary Ensemble} (GUE). The analogous scenario was first considered by Berry and Mondragon \cite{Ber87} for neutrino billiards, lacking the valley degree of freedom. When magnetic field is applied to the system, $H_{\rm Dirac}$ no longer commutes with ${\cal T}_v$ and the valley-blocks are not degenerate.  
\item[(ii)] 
In the case of {\em strong intervalley scattering} caused by irregular and abrupt system edges (or by a potential abruptly varying on the scale of atomic separation) the two sublattices are nonequivalent, so both special time-reversal symmetries ${\cal T}_{\rm sl}$ and ${\cal T}_v$ became irrelevant. For $B=|B_z|=0$, ${\cal T}$ commutes with ${\cal H}_{\rm Dirac}$ leading to the orthogonal symmetry class and statistical properties following from the {\em Gaussian Orthogonal Ensemble} (GOE) of random matrices. When increasing $B$, transition GOE-GUE similar to that discussed earlier for Schr\"{o}dinger systems \cite{Ber86} appears.
\end{enumerate}

The existing numerical studies for closed systems of irregular shapes \cite{Wur09,Der08,Lib09} show that the typical intervalley scattering time is always shorter than the time required to resolve a level spacing (Heisenberg's time) leading to the scenario (ii). The corresponding transitions between ensembles of random matrices are depicted in Fig.\ \ref{gradyn} with solid lines. Some features of the scenario (i) were found in open systems \cite{Wur09,Wur11b}, for which the intervalley scattering time needs to be compared with much shorter time characterizing the conductance (escape time). Such systems are, however, beyond the scope of this paper. We focus here on a peculiar case of {\em regular} and weakly-disordered nanosystems, for which the intervalley scattering itself may be strongly suppressed, providing the appropriate boundary conditions are chosen (see Appendix~\ref{bocodi}).

\subsection{Potential disorder in the tight-binding model on a honeycomb lattice %
   \label{dismod}}
The lattice Hamiltonian for disordered graphene in weak magnetic field reads
\begin{multline} 
  \label{hamtba}
  {\cal H}_{\rm TBA}=
  \sum_{\langle{ij}\rangle}\left[\,t_{ij}({\bf A})|{i}\rangle\langle{j}| +
    {\rm h.c.}\,\right] \\
  + \sum_i\left[\,M_V({\bf r}_i) + U_{\rm gate}({\bf r}_i) + 
    U_{\rm imp}({\bf r}_i)\,\right]|{i}\rangle\langle{i}|.
\end{multline}
The complex hopping-matrix elements are given by $t_{ij}({\bf A})=-t\exp\left[i\frac{2\pi}{\Phi_0}\int_{{\bf r}_i}^{{\bf r}_j}{\bf A}\cdot{d{\bf r}}\right]$ (with the flux quantum $\Phi_0=h/e\simeq{4.14}\times{10}^{-15}\,$T$\cdot$m$^2$) if the orbitals $|i\rangle$ and $|j\rangle$ are nearest neighbors on the honeycomb lattice (with $t=\frac{2}{3}\sqrt{3}\hbar{v}_F/a\simeq{}3\,$eV), otherwise $t_{ij}=0$. (The symbol $\sum_{\langle{ij}\rangle}$ denotes that each pair $\langle{ij}\rangle$ is counted only once.) This represents a~minimal form for the interaction between lattice fermions and the magnetic fields (Peierls construction \cite{Pei33}) in a framework of the tight-binding approximation (TBA). The mass term is modelled within a staggered potential on a~honeycomb lattice $M_V({\bf r}_i)$ \cite{Akh08}, which is positive (negative) if ${\bf r}_i$ belongs to sublattice $A$ ($B$). Typically, we put $|M_V({\bf r}_i)|\lesssim{t}$ if ${\bf r}_i$ is the outermost atom position at zigzag edge, or $M_V({\bf r}_i)={0}$ otherwise. Such a simple choice was shown to reproduce the ``infinite mass'' boundary condition correctly for various scattering problems \cite{Ryc09}. The physical origin of a staggered potential is usually related to the magnetic moments at the zigzag edges \cite{Son06}. (Alternatively, high electrochemical potential of terminal atoms can also be attributed, for instance, to the hydrogen-edge passivation, see Ref.\ \cite{Kus03}.) We also consider the case of $M_V({\bf r}_i)={0}$ at all lattice sites for the comparison. 

Finally, the electrostatic potential term in ${\cal H}_{\rm TBA}$ contains a contribution $U_{\rm gate}$ from gate electrodes (slowly varying with the site position ${\bf r}_i$) and a random contribution $U_{\rm imp}$ from impurities. For small nanoflakes one can choose $U_{\rm gate}\simeq{}U_0={\rm const}$, whereas a realization of disorder potential is generated by randomly choosing $N_{\rm imp}$ lattice sites ${\bf R}_n$ ($n=1,\dots,N_{\rm imp}$) out of $N_{\rm tot}$, and by randomly choosing the amplitudes $U_n\in(-\delta,\delta)$. The potential is then smoothed over a distance $\xi$ by convolution with a Gaussian, namely
\begin{equation} \label{uimper}
  U_{\rm imp}({\bf r})=
  \sum_{n=1}^{N_{\rm imp}}U_n\exp\left(-\frac{|{\bf r}-{\bf R}_n|^2}{2\xi^2}\right).
\end{equation}
The special case of $\xi\ll{a}$, $N_{\rm imp}=N_{\rm tot}$ corresponds to the Anderson model on a honeycomb lattice, considered in Ref.\ \cite{Der08} on spectral statistics of graphene and nanotube-like structures. Earlier, the model constituted by Eqs.\ (\ref{hamtba},\ref{uimper}) with $\xi\gg{a}$ was shown to reproduce basic transport properties of disordered mesoscopic graphene samples \cite{Ryc07,Lew08,Wur11b}. Apart from a~very recent work of Ref.\ \cite{Ryc11}, it has not been considered in the discussion of spectral statistics of nanoflakes so far.

We further define the Fourier transform of two-point correlation function
\begin{multline} 
  \label{knoddef}
  K_{\bf q}=\frac{\cal A}{\left(N_{\rm tot}\hbar{v}_F\right)^2}\sum_{i=1}^{N_{\rm tot}}
  \sum_{j=1}^{N_{\rm tot}}\left\langle
    {U_{\rm imp}({\bf r}_i)U_{\rm imp}({\bf r}_j)}\right\rangle \\
  \times\exp\left[\,i{\bf q}\cdot({\bf r}_i-{\bf r}_j)\,\right],
\end{multline}
where the system area ${\cal A}=\frac{1}{4}\sqrt{3}N_{\rm tot}a^2$, and the averaging takes place over possible realizations of the disorder (\ref{uimper}) (so $\langle{U_{\rm imp}}({\bf r})\rangle\equiv{}0$). For the length scales large compared to $\xi$, the dimensionless correlator
\begin{align} 
  \label{knodval}
  K_0 &= \frac{\sqrt{3}}{9}\,\frac{N_{\rm imp}}{N_{\rm tot}}
  \left(\frac{\delta}{t}\right)^2\kappa^2, \nonumber \\
  \kappa &= \begin{cases}
    1, & \text{if } \xi\ll{a}, \\
    \frac{8}{3}\sqrt{3}\pi(\xi/a)^2, & \text{if } \xi\gg{a},
  \end{cases}
\end{align}
becomes a representative measure of the disorder strength. For ${\bf q}\neq{0}$,
we obtain $K_{\bf q}=K_0$ if $\xi\ll{a}$, or $K_{\bf q}=K_0\exp(-q^2\xi^2)$ if $\xi\gg{a}$. (The latter justifies calling $\xi$ the 'disorder correlation length', as proposed earlier in the paper.) The numerical value of the ratio $K_{\bf q}/K_0$ at ${\bf q}=\pm{\bf K}=\left(\pm\frac{2\pi}{3a},0\right)$ roughly approximates the intervalley scattering rate \cite{Wur11b}, and is as small as $2\times{10}^{-6}$ for $\xi=\sqrt{3}\,a$ (the value used for computer simulations presented in the remaining parts of the paper). 

Apart from negligibly weak intervalley scattering discussed above, $U_{\rm imp}$ also contributes to the mass term in ${\cal H}_{\rm Dirac}$ (\ref{hameff}) and thus breaks the symplectic symmetry associated with $\tau_{\rm sl}$ (\ref{tsltv}) independently from the fact, that a similar effect may be caused by the system boundaries \cite{Luo09}. Namely, the effective mass term for low-energy excitations can be approximated by
\begin{multline}
  \label{masseff}
  M_{\rm eff}(\bar{\bf r}_{ij}) \simeq 
  \frac{1}{2}\left[M_V({\bf r}_i)-M_V({\bf r}_j)\right] \\
  + \frac{1}{2}\,\nabla{U}_{\rm imp}(\bar{\bf r}_{ij})\cdot({\bf r}_i-{\bf r}_j),
\end{multline}
where ${\bf r}_i$ and ${\bf r}_j$ are in-plane positions of atoms in the same unit cell [with $i$ ($j$) belonging to the sublattice $A$ ($B$)]. We further define $\bar{\bf r}_{ij}\equiv({\bf r}_i+{\bf r}_j)/2$. It is clear from Eq.\ (\ref{masseff}), that $M_{\rm eff}\neq{0}$ even for $M_V=0$. For higher energies, the symplectic symmetry is also broken by a nonlinear term appearing in the effective Hamiltonian derived from a nearest-neighbor tight-binding model \cite{And07}. Therefore, the structure of ${\cal H}_{\rm TBA}$ (\ref{hamtba}) provides additional reasons, for which energy levels of graphene nanoflakes in the limit of quantum chaos, obtained numerically in the remaining parts of the paper, follow GOE or GUE statistics, depending whether charge carriers are scattered between the valleys or not.

\begin{figure}
\centerline{\includegraphics[width=0.8\linewidth]{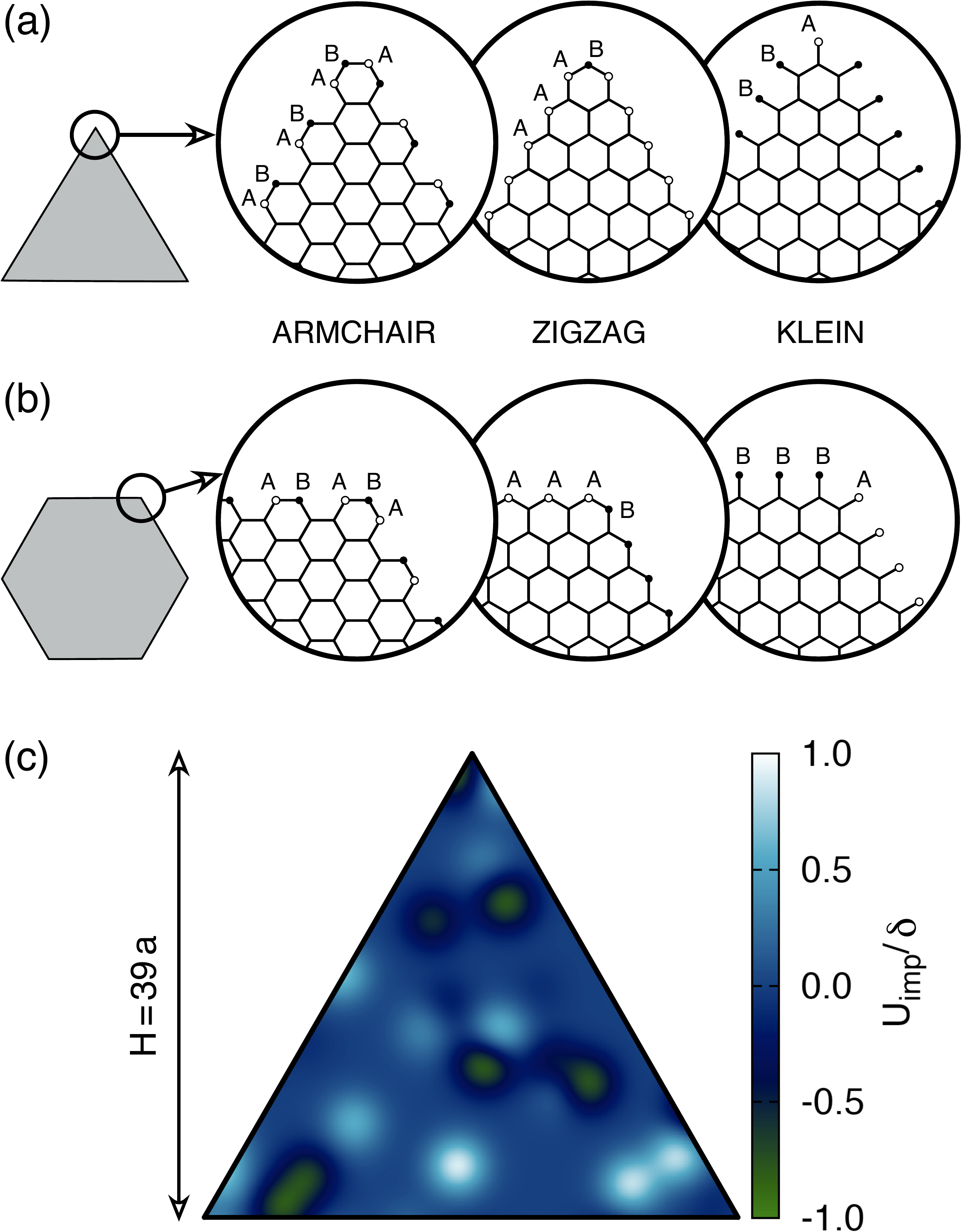}}
\caption{\label{trihex}
Systems studied numerically in the paper. (a), (b) Triangular and hexagonal nanoflakes with armchair, zigzag and Klein edges. (c) Typical impurity potential landscape $U_{\rm imp}({\bf r})$ for a triangular flake with armchair edges and 2106 carbon atoms. The triangle height is $H=39\,a\simeq{}10\,$nm, the impurity concentration and the disorder correlation length are $N_{\rm imp}/N_{\rm tot}=0.01$ and $\xi=\sqrt{3}\,a$, respectively.
}
\end{figure}

\subsection{Scope of the paper: Nanosystems considered and the numerical approach}
Graphene nanoflakes studied in the paper are shown schematically in Fig.\ \ref{trihex}. In general, we limit the discussion to the  triangular and hexagonal flakes bounded entirely with armchair, zigzag, or Klein edges [see Figs.\ \ref{trihex}(a) and \ref{trihex}(b)]. Such a choice is related to the fact, that these three types of edges were observed using different microscopy techniques \cite{Sue10,Yan10,Cai10}. In particular, STM measurements for quantum dots with well-defined edges and almost hexagonal shapes have been reported \cite{Ham11}. For this reasons, and because of numerous earlier theoretical works focused on graphene nanosystems with irregular edges \cite{Wur09,Der08,Lib09}, it is worth to find out whether quantum chaos may even appear for highly-symmetric systems in the presence weak bulk disorder, and (if so) how symmetry classes of such systems are related to the boundary conditions? Later (in Sec.\ \ref{tragogu}) we extend our analysis on flakes with some randomly-distributed edge vacancies, to make link with the results of Refs.\ \cite{Wur09,Der08,Lib09}.

\begin{table}
  \caption{ \label{sizetab}
    Types and geometric sizes of nanosystems which spectral statistics are 
    discussed in Secs.\ \ref{leretri}--\ref{tragogu} \cite{sizefoo}. 
  }
  \begin{tabular}{c|c|r|r|r|r}
    \hline\hline
    Shape & Edges & Height & \# of atoms & 
    \multicolumn{2}{|c}{Ave.\ size  $\sqrt{\cal A}$}  \\
    & & $H/a$ &  $N_{\rm tot}$ & $\ \times{a}^{-1}$ & $\ \times$nm$^{-1}$ \\ \hline
    Triangle & armchair & 78 & 8268 & 59.8 & 14.7 \\
    & & 156 & 32760 & 119.1 & 29.3 \\
    & zigzag & $45\sqrt{3}$ & 8278 & 59.9 & 14.7 \\
    & & $90\sqrt{3}$ & 32758 & 119.1 & 29.3 \\
    & Klein & $45.5\sqrt{3}$ & 8548 & 60.8 & 15.0 \\ 
    & & $90.5\sqrt{3}$ & 33298 & 120.1 & 29.5 \\ \hline
    Hexagon & armchair & 64.5 & 8322 & 60.0 & 14.8 \\
    & & 130.5 & 34062 & 121.4 & 29.9 \\
    & zigzag & $42\sqrt{3}$ & 10584 & 67.7 & 16.7 \\
    & & $84\sqrt{3}$ & 42336 & 135.4 & 33.3 \\
    & Klein & $\,41.5\sqrt{3}$ & 10332 & 66.9 & 16.5 \\
    \hline\hline
  \end{tabular}
\end{table}

As described in Sec.\ \ref{dismod}, weak bulk disorder in our nanosystems is introduced within random electrostatic potential landscape. The potential fluctuations, usually attributed to the influence of substrate impurities, give origin to the so-called ``puddles,'' i.e., spatial fluctuations in carrier density observed in numerous experiments \cite{Mar08,Zha09,Cou11}. We note here, that similar fluctuations may also arise from out-of-plane lattice deformations, modifying the electrostatic potential term \cite{Voz10}.

An example of the potential given by Eq.\ (\ref{uimper}) is shown in Fig.\ \ref{trihex}(c). For demonstrating purposes, we took a~relatively small triangular flake with armchair edges, which consists of $N_{\rm tot}=2106$ carbon atoms, corresponding to the triangle height $H=39\,a\simeq{}10\,$nm. [When analyzing statistical distributions of energy levels, we choose the systems significantly larger, see Table \ref{sizetab}.] The remaining parameters are the impurity concentration $N_{\rm imp}/N_{\rm tot}=0.01$ and the disorder correlation length $\xi=\sqrt{3}\,a$, corresponding to $K_0\simeq{}1.16\,(\delta/t)^2$. Although the impurities visualised in Fig.\ \ref{trihex}(c) are relatively well-separated from each other, as well as $K_0\ll{}1$ for a typical value of $\delta/t=0.1$ used in the simulations, we show in Secs.\ \ref{leretri} and \ref{ramaspe} that such a~weak disorder may lead to clear signatures of quantum chaos, providing the system considered is sufficiently large.

Linear sizes of nanosystems listed in Table \ref{sizetab} are close to these reported in Ref.\ \cite{Pon08}. Such systems contain $10^4\lesssim{N}_{\rm tot}<10^5$ carbon atoms, making possible to reproduce, on a~finite lattice, several features of a~continuous system (described by the Dirac theory) with a~good accuracy \cite{Two06,Ryc09}. On the other hand, the values of $N_{\rm tot}\gtrsim{}10^4$ combined with the presence of a~random potential landscape $U_{\rm imp}({\bf r})$ (\ref{uimper}), make it difficult (unless impossible) either to utilize {\em ab-initio} methods for carbon-based nanosystems \cite{Son06,Fer07,Ako08} or to employ semiclassical theory for generic Dirac billiards in uniform potentials presented in Ref.\ \cite{Wur11a}. For these reasons, our method of approach is founded on a~numerical diagonalization of tight-binding Hamiltonians ${\cal H}_{\rm TBA}$ (\ref{hamtba}) for different nanosystems and different $U_{\rm imp}({\bf r})$. In brief, when analysing the spectral statistics of smaller systems ($N_{\rm tot}\lesssim{}10^4$) we took $200-400$ randomly-chosen $U_{\rm imp}({\bf r})$ for each disorder strength quantified by the correlator $K_0$. For larger systems ($N_{\rm tot}\gg{}10^4$) we took just one $U_{\rm imp}({\bf r})$ for each $K_0$, essentially reproducing the experimental situation of Ref.\ \cite{Pon08}, where the Coulomb-blockade spectrum of a single device was obtained. This allows us to verify, whether spectral statistics obtained for the ensemble of smaller systems coincide with spectral statistics obtained for a~single, however much larger, system of each kind.

More details on our numerical approach and methods of data analysis are provided in Sec.\ \ref{ramaspe}. But first, let us briefly discuss a~level structure of integrable Dirac systems on the example of perfect triangular graphene nanoflakes, and demonstrate how the level-repulsion appears when weak disorder is included.

\section{Level repulsion in triangular graphene nanoflakes \label{leretri}}

\subsection{Energy levels of perfect triangular nanoflakes and the effect of weak disorder}
Energy levels of triangular nanoflake with armchair edges were recently found analytically, in the absence of disorder, by Rozhkov and Nori \cite{Roz10}. Close to the Dirac point, exact energies can by approximated as \cite{Ako08,Roz10}
\begin{gather}
  \label{emndiractri}
  E_{m,n}^\pm\simeq{}E_{m,n,\pm}^{\rm Dirac} = \pm
  \frac{2\pi{}t}{\sqrt{3N_{\rm tot}}}\sqrt{m^2+n^2-mn}, \\
  \label{mnatri}
  \text{with}\ \ 1\leqslant{}m\leqslant{}n.
\end{gather}
$E_{m,n,\pm}^{\rm Dirac}$ corresponds to eigenenergies of Dirac particles in a~triangular cavity with upper (lower) sign valid for electrons (holes). For zigzag edges, an analytic solution of a~tight-binding model is missing, but an approximation (\ref{emndiractri}) remains valid with $(m,n)$ obeying the conditions 
\begin{equation}
  \label{mntrig}
  m\geqslant{}1\ \ \text{and}\ \ n\geqslant{}2m.
\end{equation}
The above also applies for Klein edges.

Numerical examples of energy levels for triangular Dirac cavities and graphene nanoflakes are provided and discussed in Appendix~\ref{loentri}. Here we only mention, that for a large system vast majority of such energy levels shows the fourfold degeneracy (per each direction of spin) in the absence of disorder and for $M_V\equiv{}0$. For zigzag (or Klein) edges, this degeneracy can be easily attributed to special time-reversal symmetries ${\cal T}_{\rm sl}$ and ${\cal T}_v$ (\ref{tsltv}). For armchair edges, the symmetry associated with ${\cal T}_v$ no longer applies, but the new twofold degeneracy $E_{m,n}^{\rm Dirac}=E_{n-m,n}^{\rm Dirac}$ appears instead. This is an accidental degeneracy, which is lifted if one includes the disorder. Also, the degeneracy associated with ${\cal T}_{\rm sl}$ is lifted in the presence of disorder for all types of edges, as the disorder potential leads to the effective mass term (\ref{masseff}). In turn, only the twofold valley degeneracy (associated with ${\cal T}_v$) for zigzag or Klein edges appears to be robust against weak potential disorder. Such a degeneracy is also insesitive to the staggered potential $M_V\neq{}0$ appearing at zigzag edges.

\begin{figure}
\centerline{\includegraphics[width=\linewidth]{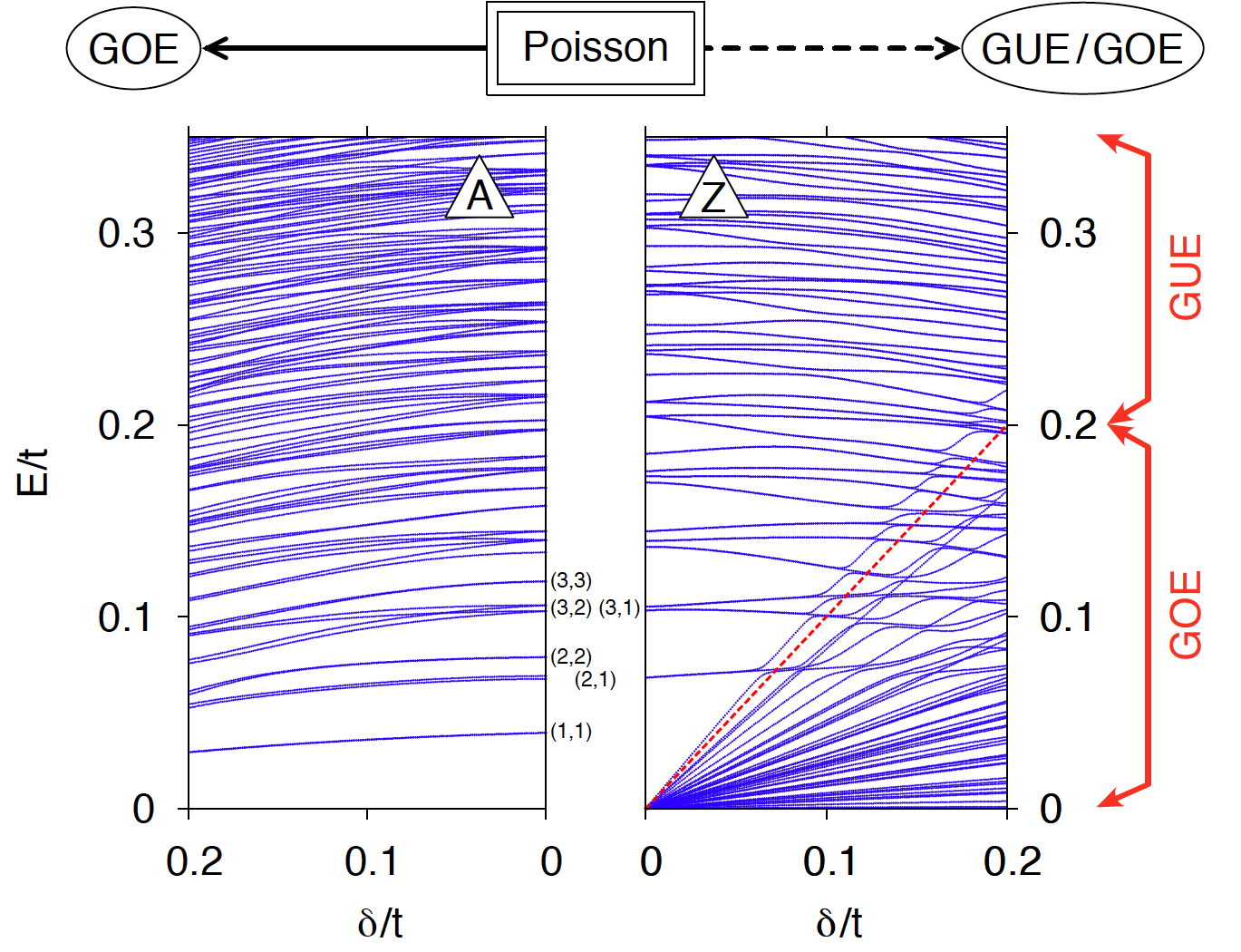}}
\caption{\label{levrep}
  Evolution of energy levels for a~triangular flake with armchair (left) and zigzag (right) edges containing $N_{\rm tot}=8268$ and $8278$ carbon atoms (respectively) when varying the disorder amplitude $\delta$ (see Ref.\ \cite{lerefoo}). The index values $(m,n)$ refers to electronic energies for Dirac cavities $E_{m,n,+}^{\rm Dirac}$ (\ref{emndiractri}). The flow-chart diagram (top) illustrates the transitions to corresponding chaotic ensembles (see Fig.\ \ref{gradyn}).
}
\end{figure}

The behavior of energy levels for triangular flakes with gradually increasing disorder is illustrated in Fig.\ \ref{levrep}. We took triangles containing $N_{\rm tot}=8268$ and $8278$ atoms (with armchair and zigzag edges, respectively), a~single disorder realization defined by positions of impurities ${\bf R}_n$ and their relative amplitudes $U_n/\delta\in(-1,1)$ for $1\leqslant{}n\leqslant{}N_{\rm imp}$ (\ref{uimper}), and varied the absolute amplitude $\delta$ \cite{lerefoo}.  Such a procedure may correspond to varying the distance between a~graphene sample and its substrate in the case of free-standing samples \cite{Gas08}, or to modifying the charge screening efectiveness in recently fabricated graphene-hBC heterostructures \cite{Bri11}.

It is clear from Fig.\ \ref{levrep}, that due to low density of states in the vicinity of the Dirac point, Eq.\ (\ref{emndiractri}) describes only the few lowest-lying states accurately for $N_{\rm tot}\lesssim{}10^4$ (notice the index values $(m,n)$ provided in the central part of a plot). The basic features of the level structure in the presence of disorder are, however, correctly reproduced. For armchair edges, degeneracies discussed above are lifted for a~relatively weak disorder strength, and one can directly compare a~level sequence obtained from the numerical diagonalization of ${\cal H}_{\rm TBA}$ (\ref{hamtba}) with predictions of the random matrix theory (RMT). For zigzag edges, the twofold (approximate) valley degeneracy of each level remains even for stronger disorder (unless $|E|\lesssim{}\delta$, see dashed line in the right panel of Fig.\ \ref{levrep}) and the additional effort in the data analysis is required. Also, when increasing the disorder strength, we observe avoided crossings signaling the level repulsion, characteristic for chaotic quantum systems \cite{Haa10}. Moreover, the repulsion is noticeably stronger for zigzag edges and $|E|>\delta$ than for the other cases. The corresponding sequence of energy levels for Klein edges is not shown in Fig.\ \ref{levrep}, but it evolves with the increasing disorder strength in an identical manner as the level sequence for zigzag edges.

These observations, together with the discussion of symmetry classes provided in Sec.\ \ref{didigra}, suggest that different ensembles of random matrices are capable of describing level sequences such as depicted in Fig.\ \ref{levrep}. Namely, GUE with the approximate valley degeneracy shall apply for triangular flakes with zigzag (or Klein) edges in the energy range $|E|>\delta$, whereas GOE (without degeneracy) shall apply otherwise. Resulting transitions between the Poissonian and different Gaussian ensembles expected when increasing the disorder strength are presented in a flow-chart diagram at the top of Fig.\ \ref{levrep}. We further verify these predictions in Sec.\ \ref{ramaspe} with the help of statistical analysis of the level sequence.

\subsection{Level-spacing distribution}
Basic statistic which distinguishes the spectra of integrable systems from chaotic ones is the level-spacing distribution $P^{(k)}(S)$. By definition, $P^{(k)}(S)dS$ represents the probability, that the quantity $\langle\rho(E)\rangle(E_{i+k}\!-\!E_i)/k$ is located in the interval $(S,S\!+\!dS)$, where $E_{i+k}\!-\!E_i$ is the distance between $k$-th neighbors in the level sequence $E_1\leqslant{}E_2\leqslant\dots$, and $\langle\rho(E)\rangle$ is the average density of levels in the energy interval $(E,E+dE)$. Unlike for Schr\"{o}dinger systems, such as the two-dimensional electron gas (2DEG) where $\langle\rho(E)\rangle$ is usually assumed to be constant, in bulk graphene
\begin{equation}
  \label{rhobulk}
  \langle\rho(E)\rangle\simeq\rho_{\rm bulk}(E)=
  \frac{1}{\pi}\frac{\cal A}{(\hbar{}v_F)^2}|E|.
\end{equation}
Due to a quasirandom character of the Hamiltonian, statistical properties of energy levels of large quantum systems are described by RMT \cite{Meh04}. For instance, generic integrable systems are described by the Poisson distribution, namely \cite{pksdef}
\begin{equation}
  \label{ppoi}
  P^{(k)}(S)=\frac{k^k}{(k-1)!}S^{k-1}\exp(-kS),
\end{equation}
with $k=1,2,\dots$. For instance, the nearest-neighbor spacing distribution $P^{(1)}(S)\equiv{}P_{\rm Poi}(S)=\exp(-S)$. In contrast, nearest-neighbor level spacings $P^{(1)}(S)$ of classically chaotic systems which preserve (or break) TRS may be approximated by the so-called Wigner surmise for GOE (or GUE) of random matrices
\begin{align}
  P_{\rm GOE}(S) &= 
  \frac{\pi}{2}S\exp\left(-\frac{\pi{}S^2}{4}\right), \label{pgoe} \\
  P_{\rm GUE}(S) &= 
  \frac{32}{\pi^2}S^2\exp\left(-\frac{4S^2}{\pi}\right). \label{pgue}
\end{align}
One notes that $P_{\rm Poi}(S)\simeq{}1-S$ for $S\rightarrow{}0$ and is maximal for $S=0$, i.e., integrable systems exhibit level attraction. In contrast, $P_{\rm GOE}(S)\propto{}S$ for $S\rightarrow{}0$ showing {\em linear} level repulsion, whereas  $P_{\rm GUE}(S)\propto{}S^2$ for $S\rightarrow{}0$ showing stronger ({\em quadratic}) level repulsion than GOE. The above holds true for a~{\em simple sequence} of energy levels \cite{Meh04}, i.e., a~sequence in which all levels have the same values of quantum numbers corresponding to strictly conserved quantities (resulting from the symmetry of the system). As discussed earlier in this Section, the valley index in graphene needs to be regarded as an example of such a~quantity for triangular nanoflakes with zigzag or Klein edges even in the presence of disorder (see also Appendix \ref{loentri}).

\begin{figure}
  \centerline{\includegraphics[width=\linewidth]{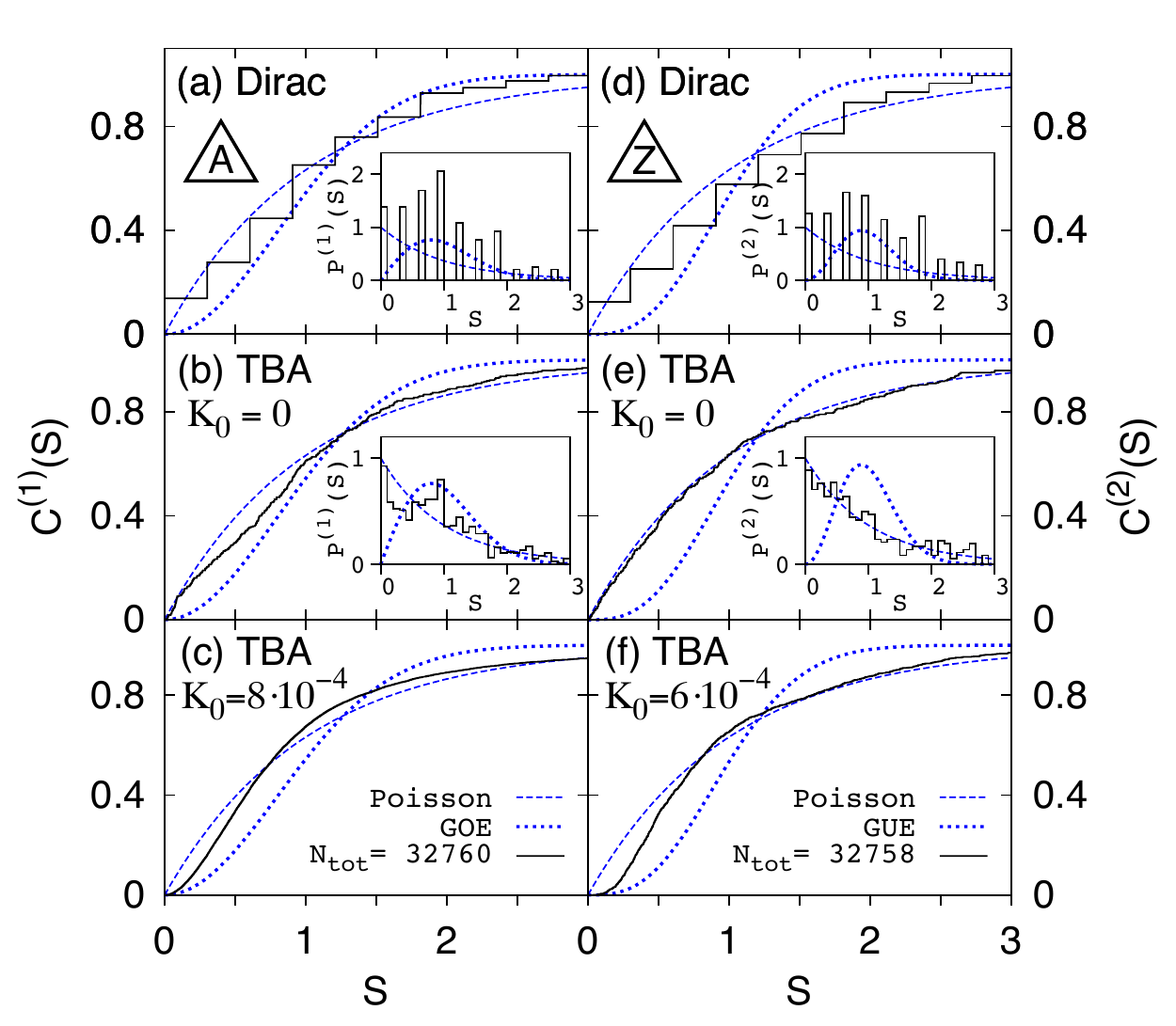}}
  \caption{\label{csklei}
    Integrated level-spacing distributions $C^{(1,2)}(S)$ for triangular flakes with armchair (a)--(c) and zigzag (d)--(f) edges, both having the area of ${\cal A}\simeq{}(120\,a)^2$ [see Table~\ref{sizetab}]. (a),(d) Levels $E_{m,n,\pm}^{\rm Dirac}$ (\ref{emndiractri}) for Dirac cavities. (b),(e) Levels obtained from diagonalization of ${\cal H}_{\rm TBA}$ (\ref{hamtba}) in the absence of disorder ($K_0\!=\!0$); and (c),(f) for infinitesimally-weak disorder ($K\!\simeq{}\!10^{-3}$) \cite{csklfoo}. Insets show level-spacing distributions $P^{(1,2)}(S)$. Numerical results are shown with solid black lines. Remaining lines are for Poisson distribution (blue dashed) and for the Wigner surmise for GOE or GUE (blue dotted).
  }
\end{figure}

Before analyzing the level-spacing distribution as a~function of disorder, we first briefly discuss that characterizing perfect (or almost perfect) triangular nanoflakes in graphene. In principle, equilateral triangles are not examples of generic integrable systems as they show number-theoretic degeneracies \cite{Ber81}. As shown in Appendix~\ref{leclutri}, the average degeneracy of levels given by Eq.\ (\ref{emndiractri}) in a~finite energy interval is divergent for $N_{\rm tot}\rightarrow\infty$ (a~phenomena known as {\em level clustering}) and $P^{(k)}(S)$ does not exists (as for similar corresponding Schr\"{o}dinger systems \cite{Pin80}). For a~large but finite $N_{\rm tot}$, the values of $S$ occurring for $E_{m,n,\pm}^{\rm Dirac}$ are equal to \cite{Agu08}
\begin{equation}
  \label{squant}
  S_q=\frac{\pi\sqrt{3}}{18}\,q\simeq 0.302\,q, \ \ \ \text{with }q=0,1,2,\dots.
\end{equation}
As a result, the integrated spacing distributions 
\begin{equation}
  \label{cksdef}
  C^{(k)}(S)=\int_0^SP^{(k)}(S')dS'
\end{equation}
show abrupt steps at $S=S_q$. 

Such steps are clearly visible in Figs.\ \ref{csklei}(a) and \ref{csklei}(d), where we plot $C^{(k)}(S)$  with $k\!=\!1$ ($k\!=\!2$) for a finite Dirac cavity with armchair (zigzag) boundary conditions [solid lines]. We took all energy levels given by Eq.\ (\ref{emndiractri}) with $|E_{m,n,\pm}^{\rm Dirac}|\leqslant{}t/2$ for $N_{\rm tot}=32760$ (or $N_{\rm tot}=32758$) and used $\rho_{\rm bulk}$ (\ref{rhobulk}) to unfold the spectrum. The statistic $C^{(2)}(S)$ is used in case of zigzag boundary conditions due to the twofold valley degeneracy of each electronic level. The steps of $C^{(k)}(S)$ are followed by peaks of $P^{(k)}(S)$ at $S=S_q$ (see insets). The discrete structure of spacing distributions gets smeared out when considering electronic levels of ${\cal H}_{\rm TBA}$ (\ref{hamtba}) even in the absence of disorder (see Figs.\ \ref{csklei}(b) and \ref{csklei}(e)). This is because ${\cal H}_{\rm TBA}$ leads to the nonlinear dispersion relation and number-theoretic degeneracies no longer apply for triangular nanoflakes (see Appendix~\ref{leclutri}). For this reason, such highly-symmetric nanosystems in graphene appear to be generic integrable systems, with Poissonian distribution of level spacings, providing the degeneracies associated with special time-reversal symmetries (\ref{tsltv}) are properly taken into account. In fact, only small deviations from $P_{\rm Poi}(S)$ [dashed lines] are visible due to finite numbers of energy levels considered (see insets in Figs.\ \ref{csklei}(b) and \ref{csklei}(e)). Finally, in Figs.\ \ref{csklei}(c) and \ref{csklei}(f) we plot $C^{(1,2)}(S)$ for triangular nanoflakes with an infinitesimally-weak bulk disorder ($K_0\!\simeq{}\!10^{-3}$) \cite{csklfoo} to illustrate the level repulsion for $S\ll{}1$. The repulsion is noticeably stronger for a~triangle with zigzag edges than for a~triangle with armchair edges, suggesting that different symmetry classes apply in these two cases (see spacing distributions obtained from the Wigner surmise for GOE and GUE; dotted lines). The evolution of spacing distributions with increasing disorder strength is analyzed in a~quantitative manner in Sec.\ \ref{ramaspe}.

\section{Random matrices and spectral statistics of disordered systems
  \label{ramaspe}}
This Section and Sec.\ \ref{tragogu} present the central results of the paper. We start from a~brief description of basic additive random-matrix models \cite{Haa10,Zyc93} capable of reproducing the evolution of spectral statistics when dynamic system undergoes transition to quantum chaos or transition between different symmetry classes. Next, we apply these models to parametrize the transition to quantum chaos in weakly-disordered graphene nanoflakes.

\subsection{Additive random-matrix models and transitions between ensembles}
When generic integrable system undergoes the transition to quantum chaos, its spectral properties may be reproduced by the following random Hamiltonian
\begin{equation}
  \label{admatm}
  H(\lambda)=\frac{H^0+\lambda{V}}{\sqrt{1+\lambda^2}},
\end{equation}
where $H^0$ is diagonal random matrix, which elements follow a Gaussian distribution with zero mean and the variance $\langle(H^0_{ij})^2\rangle=\delta_{ij}$, the parameter $\lambda\in{}[\,0,\infty\,]$, and $V=V^\dagger$ is a member of one of the Gaussian ensembles. In particular, for the transition Poisson-GOE, elements of $V$ are real numbers chosen to follow a Gaussian distribution with zero mean and the variance $\langle{V_{ij}^2}\rangle=(1+\delta_{ij})/N$, where $N$ is the matrix size. Analogously, for the transition Poisson-GUE, elements of $V$ are complex numbers which real and imaginary parts are generated independently according to Gaussian distribution with zero mean and the variance $\langle{(\mbox{Re}\,V_{ij})^2}\rangle=(1+\delta_{ij})/2N$ and  $\langle{(\mbox{Im}\,V_{ij})^2}\rangle=(1-\delta_{ij})/2N$, respectively \cite{Zyc93}.

For $N=2$, the nearest-neighbor spacing distribution for the Hamiltonian (\ref{admatm}) can be found analytically and reads, for the transition Poisson-GOE \cite{Len91},
\begin{multline}
  \label{pspoigoe}
  P_{\rm Poi-GOE}(\lambda;S)=
  \left[\frac{u(\lambda)^2S}{\lambda}\right]
  \exp\left[{-\frac{u(\lambda)^2S^2}{4\lambda^2}}\right] \\
  \times\int_0^\infty\!{d\eta}\,\exp(-\eta^2-2\lambda\eta)
  I_0\left[\frac{\eta{u(\lambda)}S}{\lambda}\right].
\end{multline}
$I_0(x)$ is the modified Bessel function of the first kind; $u(\lambda)=\sqrt{\pi}U(-\frac{1}{2},0,\lambda^2)$ with $U(a,b,x)$ the confluent hypergeometric function \cite{Abram}. For the transition Poisson-GUE we have \cite{Zyc93}
\begin{multline}
  \label{pspoigue}
  P_{\rm Poi-GUE}(\lambda;S)=
  \sqrt{\frac{2}{\pi}}\left[\frac{a(\lambda)^2S}{\lambda}\right]
  \exp\left[{-\frac{a(\lambda)^2S^2}{2\lambda^2}}\right] \\
  \times\int_0^\infty\!{d\eta}\,\frac{\exp\left(-\lambda\eta-\frac{\eta^2}{2}\right)}{\eta}\sinh\left[\frac{\eta{a(\lambda)}S}{\lambda}\right],
\end{multline}
where the coefficient $a(\lambda)$ is expressed by the error function  $\mbox{erf}(x)=(2/\sqrt{\pi})\int_0^x\exp(-t^2)dt$ as \cite{alambdapfq}
\begin{multline}
  \label{alambda}
  a(\lambda)=\sqrt{\frac{2}{\pi}}\lambda+
  \exp\left(\frac{\lambda^2}{2}\right)
  \left[1-\mbox{erf}\left(\frac{\lambda}{\sqrt{2}}\right)\right] \\ 
  + \lambda^2\int_0^\infty{d\chi}\,\frac{\exp(-\lambda{}\chi)}{\chi}
  \mbox{erf}\left(\frac{\chi}{\sqrt{2}}\right).
\end{multline}
In particular, for $\lambda=0$ Eqs.\ (\ref{pspoigoe}) and (\ref{pspoigue}) both restore the Poissonian distribution $P_{\rm Poi}(S)=\exp(-S)$. For the opposite limit ($\lambda\rightarrow\infty$) we have $P_{\rm Poi-GOE}(S)\simeq{}P_{\rm GOE}(S)$ (\ref{pgoe}) and $P_{\rm Poi-GUE}(S)\simeq{}P_{\rm GUE}(S)$ (\ref{pgue}), reproducing the Wigner surmise for GOE and GUE matrices (respectively). For $0<\lambda<\infty$, Eq.\ (\ref{pspoigoe}) describes level-spacing distributions interpolating between Poisson and GOE statistics, with $P_{\rm Poi-GOE}(\lambda;S)\propto{}S/\lambda$ if $S\lesssim\lambda\ll{1}$, or $P(\lambda;S)\propto{}S$ if $S\ll{1}\lesssim\lambda$. Analogously, Eq.\ (\ref{pspoigue}) describes level-spacing distributions interpolating between Poisson and GUE statistics, with $P_{\rm Poi-GUE}(\lambda;S)\propto{}S^2/\lambda$ if $S\lesssim\lambda\ll{1}$, or $P(\lambda;S)\propto{}S^2$ if $S\ll{1}\lesssim\lambda$. In principle, for any $\lambda>0$ the distributions (\ref{pspoigoe}) and (\ref{pspoigue}) both exhibit qualitatively the same level repulsion (i.e., linear or quadratic) as the Wigner surmise for corresponding Gaussian ensembles.

For a sake of completeness, we also mention that the random-matrix model of the form given by Eq.\ (\ref{admatm}) but describing the transition GOE-GUE (i.e., for $H^0$ a member of GOE, $V$ a member of GUE, and $N=2$) gives a~simple expression for the nearest-neighbor spacing distribution \cite{Ber86}
\begin{multline}
  \label{psgoegue}
  P_{\rm GOE-GUE}(\lambda;S)=\sqrt{\frac{2+\lambda^2}{2}}Sc^2(\lambda) \\
  \times \exp\left[-\frac{S^2c^2(\lambda)}{2}\right]\mbox{erf}
  \left[\frac{Sc(\lambda)}{\lambda}\right]
\end{multline}
with
\begin{multline}
  \label{clambda}
  c(\lambda) = \sqrt{\frac{\pi(2\!+\!\lambda^2)}{4}}\\
  \times\left[1-\frac{2}{\pi}\left(\arctan\left(\frac{\lambda}{\sqrt{2}}\right)-\frac{\sqrt{2}\lambda}{2\!+\!\lambda^2}\right)\right].
\end{multline}
By varying the parameter $\lambda\in(0,\infty)$ one gets a family of distributions interpolating between Wigner surmises $P_{\rm GOE}(S)$ (\ref{pgoe}) and $P_{\rm GUE}(S)$ (\ref{pgue}).

Despite Eqs.\ (\ref{pspoigoe}), (\ref{pspoigue}), and (\ref{psgoegue}) are exact for $2\times{}2$ random matrices only, they can also be utilized to parametrize transitions between ensembles  of large random matrices. It was show numerically, that $P_X(\lambda_{\rm fit};S)$ with $\lambda_{\rm fit}\propto\sqrt{N}\lambda$ and $X={\rm Poi-GOE}$, ${\rm Poi-GUE}$, or ${\rm GOE-GUE}$, provides approximations of the nearest-neighbor spacing distributions of random matrices given by Eq.\ (\ref{admatm}) for $N\gg{}1$ with an astonishing accuracy \cite{Zyc93}. Moreover, such approximations were applied to describe the energy spectra of various dynamic systems undergoing transitions between symmetry classes \cite{Haa10,Ber86,Zyc93,Len90}. In the remaining part of the paper we show, that distributions $P_X(\lambda_{\rm fit};S)$ are also relevant when discussing transitions between symmetry classes for Dirac fermions confined in graphene nanoflakes.

\subsection{Energy-level distributions and transition to quantum chaos %
   in graphene nanoflakes}

\begin{figure}
\centerline{\includegraphics[width=\linewidth]{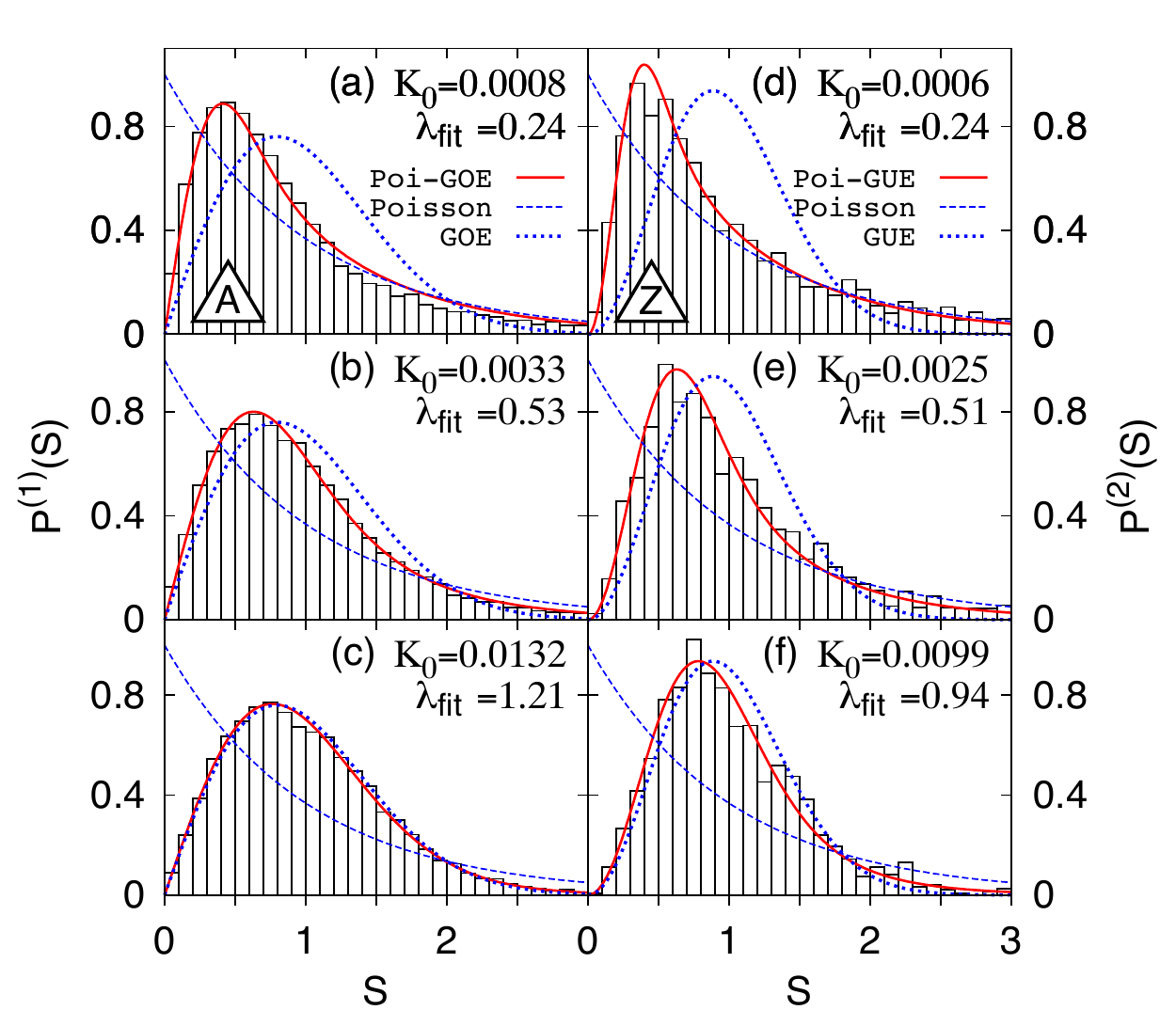}}
\caption{\label{psaztri}
  Level-spacing distributions $P^{(1,2)}(S)$ for triangular nanoflakes with armchair (a)--(c) and zigzag (d)--(f) edges. The flake area is ${\cal A}\simeq{}(120\,a)^2$. The disorder strength $K_0$ is varied between the panels \cite{psazfoo}. Numerical results are shown with black solid lines. Red solid lines show the best-fitted approximating distributions $P_{\rm Poi-GOE}(\lambda_{\rm fit};S)$ (\ref{pspoigoe}) [panels (a)--(c)] or  $P_{\rm Poi-GUE}(\lambda_{\rm fit};S)$ (\ref{pspoigue}) [panels (d)--(f)] with $\lambda_{\rm fit}$ specified for each plot. The other lines are same as in Fig.\ \ref{csklei}.
}
\end{figure}

The evolution of level-spacing distributions $P^{(1)}(S)$ (or $P^{(2)}(S)$) with the increasing disorder strength (quantified by $K_0$) is illustrated in Fig.\ \ref{psaztri} \cite{psazfoo} on the example of a~triangular nanoflake with armchair (or zigzag) edges containing $N_{\rm tot}=32760$ (or $N_{\rm tot}=32758$) atoms (see Table \ref{sizetab}). To unfold the spectra of finite systems in the presence of disorder, we use an approximating formula for the average density of states
\begin{equation}
  \label{rhoappr}
  \langle\rho(E)\rangle\simeq{}
  \rho_0 + \frac{ {\cal A}_{\rm eff} }{ {\cal A} } \rho_{\rm bulk}(E),
\end{equation}
with $\rho_{\rm bulk}(E)$ given by Eq.\ (\ref{rhobulk}). The constant term $\rho_0$ and the effective flake area ${\cal A}_{\rm eff}\lesssim{}{\cal A}$ are determined via least-square fitting of Eq.\ (\ref{rhoappr}) to the actual $\langle\rho(E)\rangle$ obtained numerically for a~particular realization of $U_{\rm imp}({\bf r})$. For the energy range considered, Eq.\ (\ref{rhoappr}) provides a~reasonable approximation of $\langle\rho(E)\rangle$ obtained for disordered graphene with various approaches, including analytical calculations employing Born approximation \cite{Hu08} or STM measurements for epitaxial graphene samples \cite{Klu09}. 

The presentation in Fig.\ \ref{psaztri} starts for infinitesimally-weak disorder $K_0\simeq{}10^{-3}$ [panels (a) and (d)], same for which integrated spacing distributions $C^{(1,2)}(S)$ are plot in Figs.\ \ref{csklei}(c) and \ref{csklei}(f). Similarly as for $C^{(1,2)}(S)$, the values of $P^{(1,2)}(S)$ [black solid lines in Fig.\ \ref{psaztri}] are intermediate between these corresponding to the Poisson distribution [blue dashed lines] and to the Wigner surmise for GOE or GUE [blue dotted lines]. The level repulsion is clearly visible at $S\ll{}1$ for both systems, but significantly stronger for the triangle with zigzag edges than for the triangle with armchair edges. Next, we enlarge the disorder strength $K_0$ by factor $4$ between the consecutive panels in Fig.\ \ref{psaztri}: (a)--(c) for armchair edges, and (d)--(f) for zigzag edges. The level-spacing distributions converge to the corresponding Wigner surmises for $K_0\gtrsim{}0.01$, see Figs.\ \ref{psaztri}(c) and \ref{psaztri}(f). The interpolating distributions $P_{\rm Poi-GOE}(\lambda_{\rm fit};S)$ (\ref{pspoigoe}) and $P_{\rm Poi-GUE}(\lambda_{\rm fit};S)$ (\ref{pspoigue}) with $\lambda_{\rm fit}$ obtained by least-square fitting [red solid lines] provide good approximations of the actual $P^{(1,2)}(S)$ for all values of $K_0$. In particular, the character of level repulsion for small $S$, which is approximately linear for transition Poisson-GOE and approximately quadratic for transition Poisson-GUE, is well-reproduced with the numerical data for triangles with armchair and zigzag edges (respectively). These are the numerical evidences showing, that symmetry classes of weakly-disordered triangular nanoflakes in graphene remain the same as predicted for chaotic Dirac billiards with appropriate boundary conditions (see Sec.\ \ref{didigra} and Appendix \ref{bocodi}), namely: the orthogonal symmetry applies for armchair edges or the unitary symmetry applies for zigzag edges. (For the latter case, the symmetry class is also insensitive to the staggered potential $M_V\neq{}0$ at the system boundary.)

\begin{figure}
\centerline{\includegraphics[width=\linewidth]{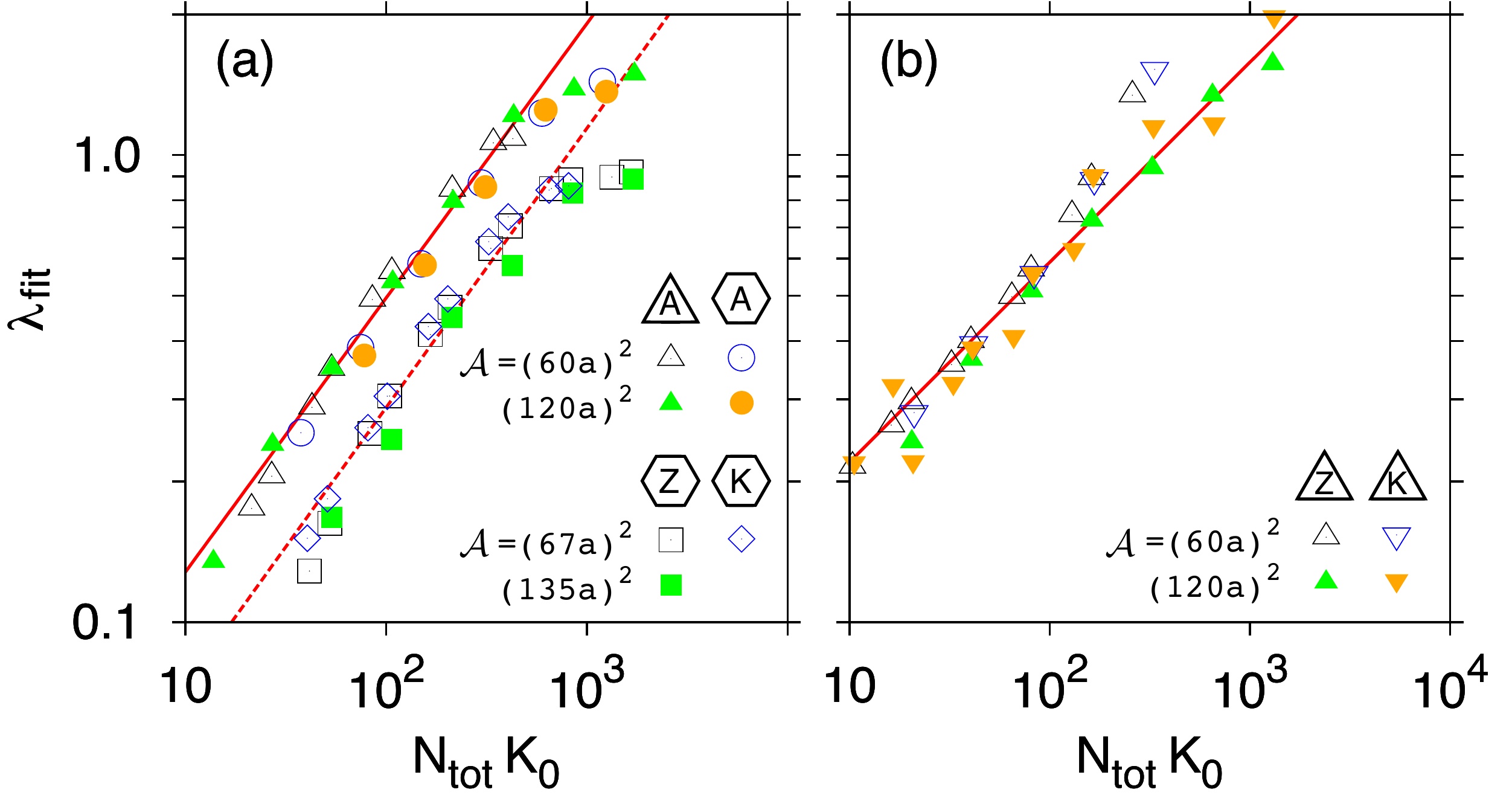}}
\caption{\label{lbfit}
  Least-squares fitted parameters $\lambda_{\rm fit}$ for (a) transition Poisson-GOE and (b) transition Poisson-GUE [see Eqs.\ (\ref{pspoigoe}) and (\ref{pspoigue}), respectively] as functions of disorder strength for systems of Table \ref{sizetab} \cite{lbfitfoo}. Datapoints in panel (a) correspond to triangles with armchair edges ({\color{black}$\triangle$} $N_{\rm tot}=8268$, {\large\color{green}$\blacktriangle$} $N_{\rm tot}=32760$) and hexagons with armchair ({\color{blue}$\bigcirc$} $N_{\rm tot}=8322$, {\Large\color{Orange}$\bullet$} $N_{\rm tot}=34062$), zigzag ({\color{black}$\square$} $N_{\rm tot}=10584$, {\color{green}$\blacksquare$} $N_{\rm tot}=42336$) and Klein ({\color{blue}$\Diamond$} $N_{\rm tot}=10332$) edges. Datapoints in panel (b) correspond to triangles with zigzag ({\color{black}$\triangle$} $N_{\rm tot}=8278$, {\large\color{green}$\blacktriangle$} $N_{\rm tot}=32758$) and Klein ({\large\color{blue}$\triangledown$} $N_{\rm tot}=8548$, {\large\color{Orange}$\blacktriangledown$} $N_{\rm tot}=33298$) edges. Logarithmic scales are used on all axes.  Lines represent the best fitted power-law relations (see Table~\ref{tab:lbfit}).
}
\end{figure}

\begin{table}
  \caption{ \label{tab:lbfit} 
    Least-square fitted parameters in Eq.\ (\ref{lbzeta}) corresponding to lines in Figs.\ \ref{lbfit}(a) and \ref{lbfit}(b). The values for hexagons with Anderson-type disorder ($\xi=0$, $N_{\rm imp}=N_{\rm tot}$) are provided in the last row for comparison. Numbers in parenthesis are standard deviations for the last digit. }
  \begin{tabular}{c|c|c|c}
    \hline\hline
    Shape & Edges & $\lambda_1$ & $\alpha$  \\ \hline
    Triangle/hexagon & armchair & $\,0.033(4)^a$ & $\,0.59(3)^a$  \\ 
    Hexagon & zigzag/Klein & $\,0.018(3)^a$ & $\,0.60(3)^a$  \\ 
    Triangle & zigzag/Klein & $\,0.082(8)^b$ & $\,0.43(2)^b$  \\ \hline
    Hexagon, $\xi=0$ & zigzag & $\,0.046(3)^a$ & $0.59(1)^a$ \\
    \hline\hline
  \end{tabular}\\
  $^a$Transition Poisson-GOE. $^b$Transition Poisson-GUE.
\end{table}

Similar agreement between $P_{\rm Poi-GOE}(\lambda_{\rm fit};S)$ (or $P_{\rm Poi-GUE}(\lambda_{\rm fit};S)$) and the actual level-spacing distributions was observed for all nanosystems listed in Table \ref{sizetab}. In Fig.\ \ref{lbfit} we plot the values of $\lambda_{\rm fit}$ as functions of the {\em total} disorder strength $N_{\rm tot}K_0$ \cite{lbfitfoo}. We find such an extensive quantity makes it possible to identify the approximating power-law relations 
\begin{equation}
  \label{lbzeta}
  \lambda_{\rm fit}\simeq\overline{\lambda}_{\rm fit}(\zeta)=
  \lambda_1\zeta^\alpha
  \ \ \ \text{with}\ \ \ \zeta\equiv{}N_{\rm tot}K_0.
\end{equation}
The coefficients $\lambda_1$ and $\alpha$ (provided in Table~\ref{tab:lbfit}) still differ between the systems with different shapes or boundary conditions, but remain unchanged when varying $N_{\rm tot}$ and $K_0$ independently with the remaining parameters fixed. Surprisingly, the datapoints obtained for the entire collection of quantum billiards listed in Table \ref{sizetab} group along just three distinct lines on the log-log plot (see Fig.\ \ref{lbfit}). The datapoints corresponding to either triangles or hexagons with armchair edges (showing the transition Poisson-GOE when increasing the disorder strength) may be approximated by $\overline{\lambda}_{\rm fit}(\zeta)$ (\ref{lbzeta}) with the parameters $\lambda_1$ and $\alpha$ given in the first row of Table~\ref{tab:lbfit} [red solid line in Fig.\ \ref{lbfit}(a)]. The datapoints corresponding to hexagons with zigzag or Klein edges (also showing the transition Poisson-GOE) may be approximated by $\overline{\lambda}_{\rm fit}(\zeta)$ (\ref{lbzeta}) with $\lambda_1$ and $\alpha$ given in the second row of Table~\ref{tab:lbfit} [red dashed line in Fig.\ \ref{lbfit}(a)]. We notice, that the best-fitted values of the exponent $\alpha$ for these two situations equal to $\alpha_{\rm Poi-GOE}\simeq{}0.6$ within the range of errorbars. Also, the same value of $\alpha$ was obtained for hexagons with Anderson-type disorder (see the last row in Table~\ref{tab:lbfit}), suggesting that it is specific for nanosystems undergoing the transition Poisson-GOE, regardless the microscopic details of the disorder realization. Finally, for triangles with zigzag or Klein edges (showing the transition Poisson-GUE) the corresponding datapoints shown in Fig.\ \ref{lbfit}(b) may be approximated by $\overline{\lambda}_{\rm fit}(\zeta)$ (\ref{lbzeta}) with parameters given in the third row of Table~\ref{tab:lbfit} [red solid line]. The exponent $\alpha$ is equal to $\alpha_{\rm Poi-GUE}\simeq{}0.4$ in this case.
 
The numerical results presented in Figs.\ \ref{psaztri} and \ref{lbfit} constitute an onset of transition to quantum chaos in highly-symmetric graphene nanoflakes with a~weak potential disorder. The actual level-spacing distributions follow $P_{\rm Poi-GOE}(\lambda;S)$ (\ref{pspoigoe}) or $P_{\rm Poi-GUE}(\lambda;S)$ (\ref{pspoigue}) obtained from basic random-matrix models, which are applicable for generic quantum system in the orthogonal or the unitary symmetry class. Depending whether the intervalley scattering is strong or weak in a particular graphene nanosystem, we have $P^{(1)}(S)\simeq{}P_{\rm Poi-GOE}(\lambda_{\rm fit};S)$ or $P^{(2)}(S)\simeq{}P_{\rm Poi-GUE}(\lambda_{\rm fit};S)$ (the systems conserving valley pseudospin show an approximate twofold degeneracy of each energy level even in the presence of disorder). The parameter $\lambda_{\rm fit}\propto{}(N_{\rm tot}K_0)^\alpha$, with the exponent $\alpha$ taking one of the two values: $\alpha_{\rm Poi-GOE}\simeq{}0.6$, or $\alpha_{\rm Poi-GUE}\simeq{}0.4$. These characteristics of transition to quantum chaos are further supported by the behavior of more distant spacings distributions briefly discussed in the next subsection.

\subsection{Spectral rigidity}
A customary measure of spectral fluctuations on scales longer than spacings described by $P^{(1)}(S)$ or $P^{(2)}(S)$ is provided by the {\em spectral rigidity} $\Delta_3(L)$ defined by Dyson and Mehta \cite{Dys63}
\begin{equation}
  \label{del3def}
  \Delta_3(L)=\frac{1}{L}\left<\underset{\scriptsize (a,b)}{\mbox{Min}}
    \int_{-L/2}^{L/2}dx\left[{\cal N}(x_0+x)-ax-b\right]^2
  \right>,
\end{equation}
where $x\equiv\langle{\cal N}(E)\rangle$ and ${\cal N}(E)$ denotes the number of energy levels having energy between $E_{\rm min}>0$ and $E_{\rm max}\geqslant{}E\geqslant{}E_{\rm min}$. In turn, the average $\langle{\cal N}(E)\rangle=\int_{E_{\rm min}}^EdE\langle\rho(E)\rangle$ with $\langle\rho(E)\rangle$ approximated by Eq.\ (\ref{rhoappr}). [For negative $E$, $\langle{\cal N}(E)\rangle=-\int_E^{-E_{\rm min}}dE\langle\rho(E)\rangle$.] The spectral rigidity $\Delta_3(L)$ represents the-least square deviation of the actual spectral staircase ${\cal}N(E)$ from the best-fitting function $ax+b$ over a range $x\in{}(x_0-L/2,x_0+L/2)$ (the averaging in Eq.\ (\ref{del3def}) runs over the interval center $x_0$). Theoretical expectations for $\Delta_3(L)$ are \cite{MehA40,Dys63}
\begin{widetext}
\begin{gather}
  \Delta_3^{(X)}(L)=
  \frac{L}{15} -\frac{1}{15L^4}\int_0^LdS(L-S)^3(2L^2-9LS-3S^2)Y_X(S), 
  \ \ \ \ 
  \text{with } X={\rm Poi}, {\rm GOE}, {\rm GUE}; \label{del3th} \\
  Y_{\rm Poi}(S)=0, 
  \ \ \ \
  Y_{\rm GOE}(S) = \left[\frac{\sin(\pi{}S)}{\pi{}S}\right]^2
  +\frac{d}{dS}\left[\frac{\sin{}(\pi{}S)}{\pi{}S}\right] 
  \int_S^{\infty}\frac{ \sin{}(\pi{}t) }{ \pi{}t }dt, 
  \ \ \ \
  Y_{\rm GUE}(S) = \left[\frac{\sin(\pi{}S)}{\pi{S}}\right]^2. 
  \label{ysgue}  
\end{gather}
\end{widetext}
In particular, for the Poisson distribution we have $\Delta_3^{(\rm Poi)}(L)=\frac{1}{15}L$. For $X={\rm GOE}$ or GUE, $\Delta_3^{(X)}(L)\simeq{}\frac{1}{15}L$ for $L\ll{}1$ \cite{del3foo}, and the exact values for larger $L$ can be obtained numerically from Eqs.\ (\ref{del3th},\ref{ysgue}).

\begin{figure}
\centerline{\includegraphics[width=0.9\linewidth]{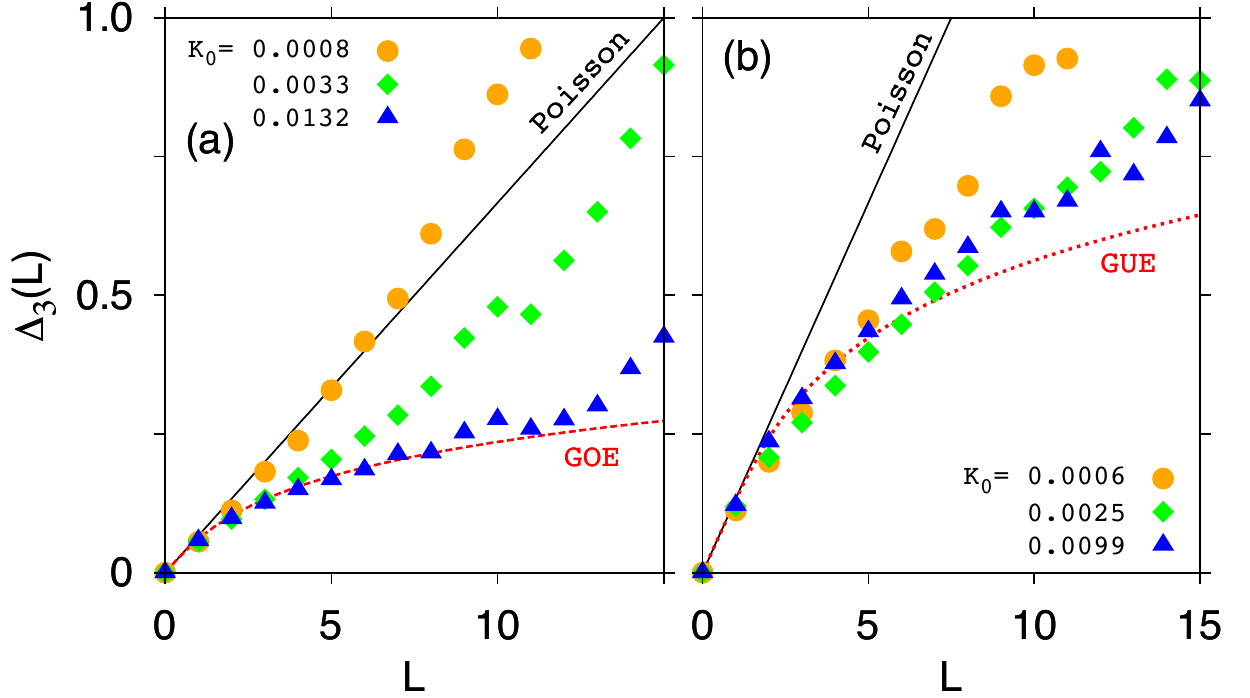}}
\caption{\label{del3az}
  Spectral rigidity $\Delta_3(L)$ for triangular nanoflakes with armchair (a) and zigzag (b) edges. Datapoints show the results obtained numerically using the same system parameters as in Fig.\ \ref{psaztri}. Lines correspond to the theoretical expectations [see Eqs.\ (\ref{del3th},\ref{ysgue})] for Poisson distribution (black solid), GOE (red dashed), and GUE (red dotted). (Notice that theoretical curves on panel (b) are rescaled via $\tilde{\Delta}_3^{(X)}(L)=4\Delta_3^{(X)}(L/2)$ due to the twofold valley degeneracy of each energy level.)
}
\end{figure}

Figs.\ \ref{del3az}(a) and \ref{del3az}(b) show the spectral rigidity $\Delta_3(L)$ for graphene nanoflakes same as in Figs.\ \ref{psaztri}(a) and \ref{psaztri}(b) (see also Ref.\ \cite{psazfoo}). It is clear from Fig.\ \ref{del3az}(a), that the spectral rigidity for triangles with armchair edges gradually evolves, with increasing $K_0$, from the straight line for Poisson distribution (black solid line) to the curve depicting the theoretical prediction for GOE (red dashed line) obtained from Eqs.\ (\ref{del3th},\ref{ysgue}). Some deviations visible for $L\gtrsim{}10$ can be attributed to a finite system size (see also the second paper in Ref.\ \cite{Lib09}). For triangles with zigzag edges, the approximate twofold valley degeneracy is observed, and the theoretical predictions $\Delta_3^{(X)}(L)$ need to be replaced by $\tilde{\Delta}_3(L)=4\Delta_3^{(X)}(L/2)$, drawn in Fig.\ \ref{del3az}(b) for $X={\rm Poi}$ (black solid line) and $X={\rm GUE}$ (red dotted line). The evolution of the actual $\Delta_3(L)$ between the limiting theoretical curves is observed also in these case, showing the convergence to the predictions for GUE is reached, in the range $L\lesssim{}10$, for as small disorder strengths as $K_0\simeq{}10^{-2}$.

\section{Transition GOE-GUE in triangular graphene nanoflakes %
   \label{tragogu}}

In this Section, we first discuss the transition GOE-GUE at zero magnetic field on the example of a triangular graphene nanoflake with zigzag edges, a finite number of edge vacancies $N_{\rm vac}$ (which modifies the intervalley scattering rate), and the bulk disorder strong enough to drive the system into chaotic regime. Then, we demonstrate the evolution of spectral statistics with the increasing magnetic flux $\Phi$ piercing the system.

\subsection{Level-spacing distributions in the presence of edge vacancies %
  \label{goguvac} }

\begin{figure}
\centerline{\includegraphics[width=\linewidth]{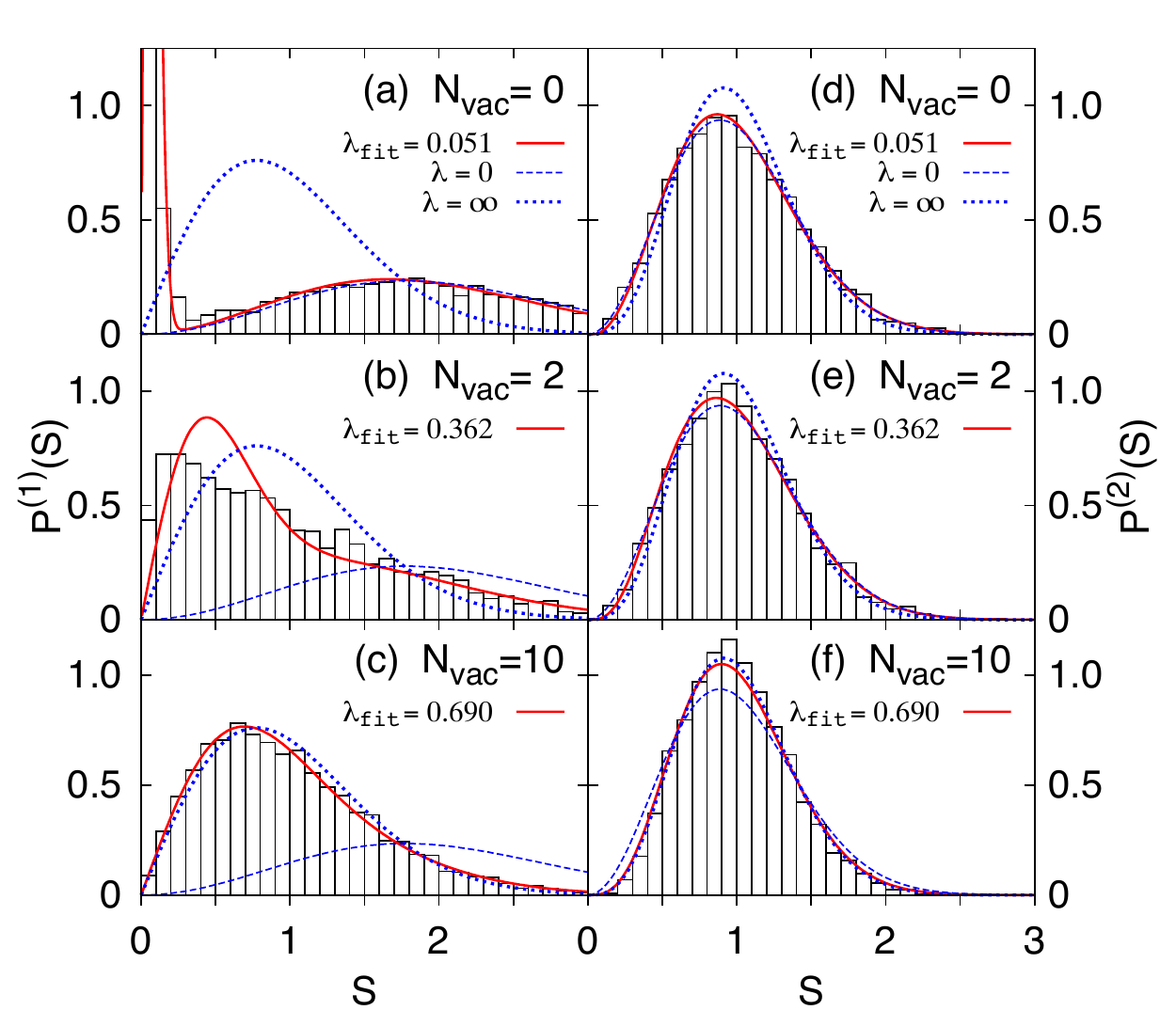}}
\caption{\label{psnvac}
  Level-spacing distributions $P^{(1)}(S)$ (a)--(c) and $P^{(2)}(S)$ (d)--(f) for triangular nanoflakes with zigzag edges, the area ${\cal A}=(120\,a)^2$, and the disorder strength fixed at $K_0\simeq{}0.04$ (corresponding to $N_{\rm tot}K_0\simeq{}1.3\times{}10^3$) \cite{psnvfoo}. The number of edge vacancies $N_{\rm vac}$ is varied between the panels. Numerical results are shown with black solid lines. The other lines correspond to empirical distributions $P_{\overline\alpha,\overline\kappa}^{(1)}(\lambda;S)$ (\ref{ps1barak}) [or $P_{\overline\alpha,\overline\kappa}^{(2)}(\lambda;S)$ (\ref{ps2barak})] with least-square fitted $\lambda=\lambda_{\rm fit}$ (red solid), $\lambda=0$ (blue dashed), and $\lambda=\infty$ (blue dotted).
}
\end{figure}

\begin{figure}
\centerline{\includegraphics[width=\linewidth]{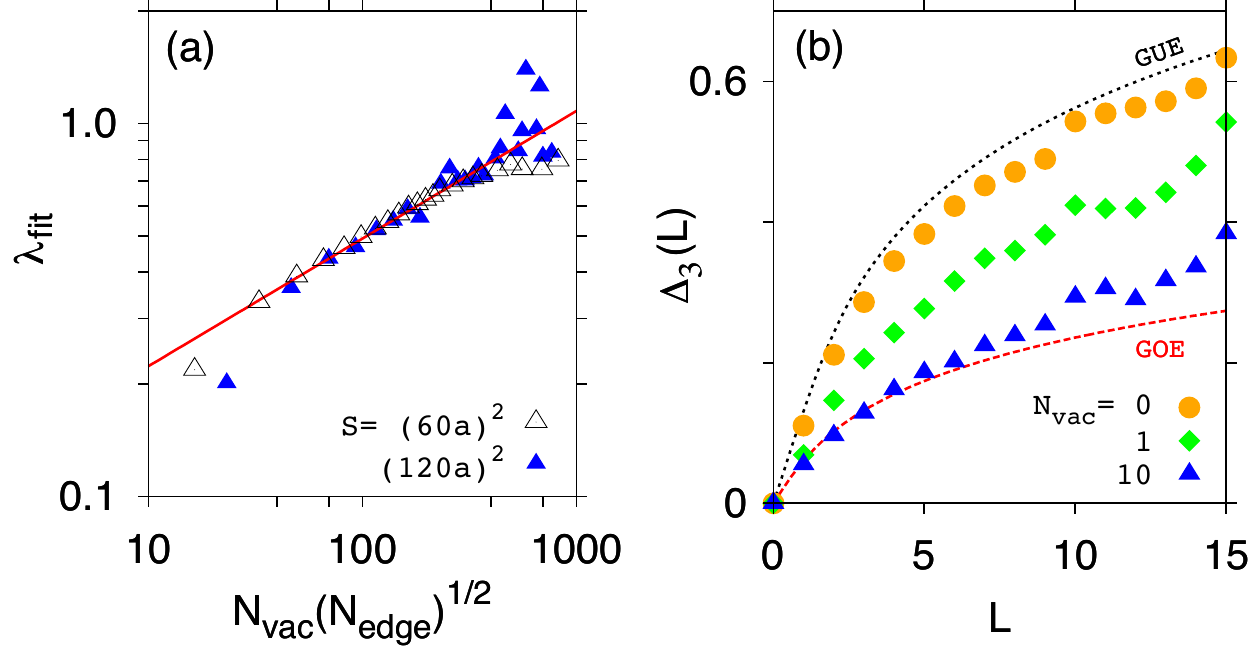}}
\caption{\label{lbnvac}
  (a) Least-squares fitted parameters $\lambda_{\rm fit}$ of empirical distributions $P_{\overline\alpha,\overline\kappa}^{(1)}(\lambda_{\rm fit};S)$ (\ref{ps1barak}) approximating actual nearest-neighbor spacing distributions $P^{(1)}(S)$ for triangular nanoflakes with zigzag edges and finite number of edge vacancies $N_{\rm vac}$.
Flake areas are ${\cal A}\simeq{}(60\,a)^2$ (open symbols) and $(120\,a)^2$ (close symbols), corresponding to the total number of terminal sites $N_{\rm edge}=270$ and $540$ (respectively). Solid line depicts the best-fitted power-law relation (\ref{lbpower}). (b) Spectral rigidity $\Delta_3(L)$ obtained numerically for ${\cal A}=(120\,a)^2$ with $N_{\rm vac}=0$ (circles), $N_{\rm vac}=1$ (diamonds), and $N_{\rm vac}=10$ (triangles). Lines correspond to theoretical expectations (\ref{del3th}) for GOE (red dashed) and for GUE with the twofold (valley) degeneracy of each level (black dotted). The disorder strength is fixed at $K_0\simeq{}0.04$ \cite{psnvfoo} for all systems. 
}
\end{figure}

Spectral characteristics of triangular nanoflakes with zigzag edges and a~finite number ($N_{\rm vac}$) of vacancies, randomly-distributed at the system boundary, are presented in Figs.\ \ref{psnvac} and \ref{lbnvac}. In particular, the evolution of nearest-neighbor spacing distribution $P^{(1)}(S)$ for $0\leqslant{}N_{\rm vac}\leqslant{}10$ is shown in Figs.\ \ref{psnvac}(a)--(c), where we have chosen the total disorder strength $N_{\rm tot}K_0\simeq{}1.3\times{}10^3$ \cite{psnvfoo}. 

Earlier, we found that second-neighbor spacing distribution $P^{(2)}(S)$  for the system without vacancies ($N_{\rm vac}=0$) can be approximated by $P_{\rm Poi-GUE}(\lambda_{\rm fit},S)$ (\ref{pspoigue}) with $\lambda_{\rm fit}=1.62$ (see Fig.\ \ref{lbfit}(b)), truncating the distribution following from GUE of random matrices with approximate twofold (valley) degeneracy of each level. This observation is further supported by the bimodal structure of $P^{(1)}(S)$ visible in Fig.\ \ref{psnvac}(a). The first mode of the distribution obtained numerically using levels with energies $0.1\leqslant{}|E|/t\leqslant{}0.5$ (black solid line) is centered at $S\simeq{}0$, and corresponds to the contribution of {\em odd} spacings, separating two almost-degenerated copies of each energy level belonging to $K$ and $K'$ valleys. (Notice, that weak intervalley scattering is present in a~finite lattice system even for $N_{\rm vac}=0$.) The second mode, centered at $S\simeq{}2$, corresponds to the contribution of {\em even} spacings, and reproduces the Wigner surmise for GUE (\ref{pgue}) scaled according to $P^{(1)}(S)\simeq\frac{1}{4}P_{\rm GUE}(S/2)$ (blue dashed line) with an excellent accuracy for $S\gtrsim{}1$. For $N_{\rm vac}>0$, the contribution of odd spacings gets smeared out and the distribution $P^{(1)}(S)$ follows the corresponding GOE statistic (blue dotted line) starting from moderate numbers of edge vacancies (see Figs.\ \ref{psnvac}(b) for $N_{\rm vac}=2$ and \ref{psnvac}(c) for $N_{\rm vac}=10$).

The features of $P^{(1)}(S)$ presented above can be rationalized with the ansatz for odd and even spacings
\begin{align}
  P_{\rm odd}(\alpha;S) &= \alpha{}P_{\rm GOE}(\alpha{}S), 
  \label{psodd} \\
  P_{\rm even}(\beta,\kappa;S) &= \beta{}P_{\rm GOE-GUE}(\kappa;\beta{}S), 
  \label{pseven}
\end{align}
with $P_{\rm GOE}(S)$, $P_{\rm GOE-GUE}(\lambda;S)$ given by Eqs.\ (\ref{pgoe},\ref{psgoegue}) and a~constrain $\beta=\alpha/(2\alpha-1)$ guaranteeing, that the resulting distribution $\frac{1}{2}[P_{\rm odd}+P_{\rm even}]$ satisfies $\langle{}S\rangle=1$ for $1\leqslant{}\alpha\leqslant\infty$ and $0\leqslant{}\kappa\leqslant\infty$. In Appendix~\ref{tra4mat}, we propose real random-matrix model with a~single parameter $\lambda$ controlling the transition from GUE with twofold level degeneracy ($\lambda=0$) to GOE without the degeneracy ($\lambda=\infty$) and utilize the ansatz (\ref{psodd},\ref{pseven}). The empirical relations $\alpha=\overline\alpha(\lambda)$ (\ref{alpemp}) and $\kappa=\overline\kappa(\lambda)$ (\ref{kapemp}) allow us to consider an approximating formula for nearest-neighbor spacing distributions 
\begin{equation}
  \label{ps1barak}
  P_{\overline\alpha,\overline\kappa}^{(1)}(\lambda;S)=\frac{
    P_{\rm odd}(\overline\alpha(\lambda);S) + 
    P_{\rm even}(\overline\beta(\lambda),\overline\kappa(\lambda);S)
  }{2}.
\end{equation}
For second-neighbor spacing distributions, we take
\begin{multline}
  \label{ps2barak}
  P_{\overline\alpha,\overline\kappa}^{(2)}(\lambda;S)=2\int_0^{2S}dS'
  P_{\rm odd}(\overline\alpha(\lambda);2S-S') \\
  \times
  P_{\rm even}(\overline\beta(\lambda),\overline\kappa(\lambda);S').
\end{multline}

The empirical distributions $P_{\overline\alpha,\overline\kappa}^{(1)}(\lambda;S)$ and $P_{\overline\alpha,\overline\kappa}^{(2)}(\lambda;S)$ with the parameter $\lambda=\lambda_{\rm fit}$ (best-fitted for each value of $N_{\rm vac}$) are shown in Fig.\ \ref{psnvac} with red solid lines. The asymptotic forms of $P_{\overline\alpha,\overline\kappa}^{(1,2)}(\lambda;S)$ for $\lambda=0$ and $\lambda=\infty$ are depicted with blue dashed and blue dotted lines (respectively). In the first limit ($\lambda=0$) we have 
\begin{align}
  P_{\overline\alpha,\overline\kappa}^{(1)}(0;S) &= 
  \frac{1}{2}\delta(S)+\frac{1}{4}P_{\rm GUE}(S/2), \label{psak1zero} \\
  P_{\overline\alpha,\overline\kappa}^{(2)}(0;S) &= P_{\rm GUE}(S), \label{psak2zero}
\end{align}
restoring spectral properties of GUE with the exact twofold degeneracy of each level. The actual spacing distributions for $N_{\rm vac}=0$ (see Fig.\ \ref{psnvac}(a) for $P^{(1)}(S)$ and \ref{psnvac}(d) for $P^{(2)}(S)$; black solid lines) show small deviations from $P_{\overline\alpha,\overline\kappa}^{(1,2)}(0;S)$ and can be approximated by $P_{\overline\alpha,\overline\kappa}^{(1,2)}(\lambda_{\rm fit};S)$ with $\lambda_{\rm fit}=0.051$, providing an estimation of the intervalley scattering rate \cite{lamzefoo} in a~triangular nanoflake with perfect zigzag edges. In the opposite limit ($\lambda=\infty$) \cite{laminfoo} 
\begin{align}
  P_{\overline\alpha,\overline\kappa}^{(1)}(\infty;S) & \simeq{} P_{\rm GOE}(S), \\
  P_{\overline\alpha,\overline\kappa}^{(2)}(\infty;S) & \simeq{} 
  2\int_0^{2S}dS'P_{\rm GOE}(2S-S')P_{\rm GOE}(S') \nonumber \\ 
  & = \pi\exp\left(-\pi{}S^2\right) \left[
  S+\frac{\pi{}S^2-1}{\sqrt{2}} \right. \nonumber\\ 
  & \ \ \ \left.\times\exp\left(\frac{\pi{}S^2}{2}\right)\mbox{erf}\left(\sqrt{\frac{\pi}{2}}S\right)\right].
\end{align}
The spacing distributions $P^{(1,2)}(S)$ for $N_{\rm vac}=10$ [see Figs.\ \ref{psnvac}(c) and \ref{psnvac}(f)] are close to $P_{\overline\alpha,\overline\kappa}^{(1,2)}(\infty;S)$, but they still fit to the approximating distributions $P_{\overline\alpha,\overline\kappa}^{(1,2)}(\lambda_{\rm fit};S)$ with $\lambda_{\rm fit}=0.69$ noticeably better. 

For the intermediate values of $N_{\rm vac}$, we generally observe some systematic deviations of $P^{(1)}(S)$ from the approximating distribution $P_{\overline\alpha,\overline\kappa}^{(1)}(\lambda_{\rm fit};S)$ for $S\lesssim{}1$ and good agreement for larger $S$. However, the comparison of $P^{(1)}(S)$ shown in Fig.\ \ref{psnvac}(b) with the corresponding distribution for smaller system [with the area ${\cal A}\simeq{}(60\,a)^2$] suggests that $P^{(1)}(S)$ gradually converges to  $P_{\overline\alpha,\overline\kappa}^{(1)}(\lambda_{\rm fit};S)$ with the system size also for $S\lesssim{}1$. No significant deviations of $P^{(2)}(S)$ from  $P_{\overline\alpha,\overline\kappa}^{(2)}(\lambda_{\rm fit};S)$ are observed.

In Fig.\ \ref{lbnvac}(a) we plot the values of $\lambda_{\rm fit}$ for two different flake areas ${\cal A}\simeq{}(60\,a)^2$ and $(120\,a)^2$ (open and close symbols, respectively) as functions of the variable $N_{\rm vac}N_{\rm edge}^{1/2}$, where $N_{\rm edge}$ denotes the total number of terminal sites. Such a~scaling allows us to find a single power-law relation for systems of different sizes, namely 
\begin{equation}
  \label{lbpower}
  \lambda_{\rm fit}\simeq{}0.102(4)\times{}\left[N_{\rm vac}N_{\rm edge}^{1/2}\right]^{0.34(1)},
\end{equation}
with the numerical values of parameters obtained via least-squares fitting (the standard deviation of a last digit are specified by numbers in parenthesis). The approximating relation given by Eq.\ (\ref{lbpower}) is also depicted in Fig.\ \ref{lbnvac}(a) [red solid line].

The evolution of more distant level spacings with increasing $N_{\rm vac}$ is illustrates in Fig.\ \ref{lbnvac}(b), where we plot the spectral rigidity $\Delta_3(L)$ for $0\leqslant{}N_{\rm vac}\leqslant{}10$. For $N_{\rm vac}=0$, $\Delta_3(L)$ obtained numerically closely follows the theoretical expectation for GUE with the twofold valley degeneracy of each level $\tilde{\Delta}_3^{\rm (GUE)}(L)=4\Delta_3^{\rm (GUE)}(L/2)$ (\ref{del3th}) (see circles and black dotted line, respectively). When increasing $N_{\rm vac}$, the datapoints (diamonds or triangles for $N_{\rm vac}=1$ or $10$) gradually approaches $\Delta_3^{\rm (GOE)}(L)$ (red dashed line) for all values of $L\lesssim{}10$. (With some deviations for larger $L$ due to a finite system size.)

Our demonstration of the nonstandard transition GUE-GOE in graphene nanoflakes, driven by varying the intervalley scattering rate at zero magnetic field, is now complete. Most remarkably, basic spectral characteristics start to reproduce those obtained in Refs.\ \cite{Wur09,Der08,Lib09} for irregular edges, after removing just a~few percent of terminal atoms from the system with perfect zigzag edges.

\subsection{Influence of external magnetic fields \label{gogufi}}

\begin{figure}
\centerline{\includegraphics[width=\linewidth]{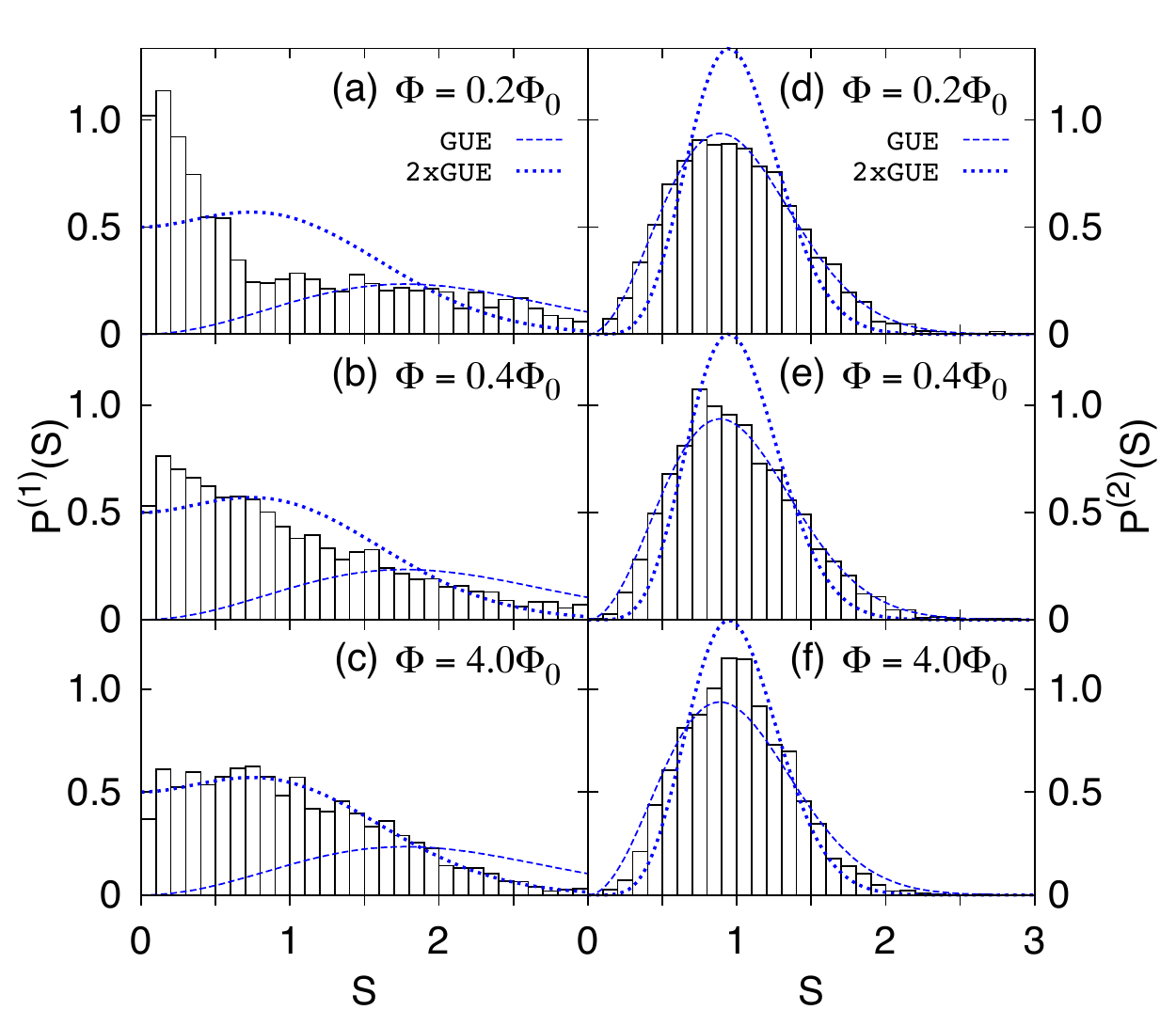}}
\caption{\label{psfi}
  Level-spacing distributions $P^{(1)}(S)$ (a)--(c) and $P^{(2)}(S)$ (d)--(f) for the same system as in Fig.\ \ref{psnvac} (with $N_{\rm vac}=0$) placed in the uniform magnetic field (with the total flux $\Phi$ specified at each panel). Numerical results are shown with black solid lines. Other lines are for GUE with twofold degeneracy of each level, see Eqs.\ (\ref{psak1zero},\ref{psak2zero}) [blue dashed]; and for two independent GUEs, see Eqs.\ (\ref{p2xgue1},\ref{p2xgue2}) [blue dotted].
}
\end{figure}

At finite magnetic fields, spectral statistics of graphene nanosystems with negligibly-weak intervalley scattering (a~situation occurring for triangular nanoflakes with zigzag edges and $N_{\rm vac}=0$) also require some attention. In Fig.\ \ref{psfi} we plot level-spacing distributions for the flake area ${\cal A}\simeq{}(120\,a)^2$, and different magnetic fields $B$ (quantified by the total flux piercing the system area $\Phi={\cal A}B$). The actual spacing distributions  $P^{(1)}(S)$ and $P^{(2)}(S)$ (black solid lines), obtained numerically for the remaining system parameters same as in Fig.\ \ref{psnvac}, show crossovers from the theoretical predictions for GUE with twofold valley degeneracy (blue dashed lines) given by Eqs.\ (\ref{psak1zero},\ref{psak2zero}) to the predictions for a level sequence following from two statistically-independent GUEs (blue dotted lines), approaching the latter for $\Phi\gtrsim\Phi_0$. For such a combined sequence, we have the distribution of a Berry-Robnik type \cite{Ber84}
\begin{align}
  P^{(1)}_{2\times\text{GUE}}(S) &= \frac{d^2}{dS^2}\left[E_{\rm GUE}(S/2)\right]^2, 
  \label{p2xgue1} \\
  P^{(2)}_{2\times\text{GUE}}(S) &= 
  2\int_0^{2S}\,dS'P^{(1)}_{2\times\text{GUE}}(2S\!-\!S') 
  P^{(1)}_{2\times\text{GUE}}(S'), \label{p2xgue2}
\end{align}
where $E_{\rm GUE}(S)= e^{-4S^2/\pi} - S + S\,\mbox{erf}\left(2S/\sqrt{\pi}\right)$ \cite{eguefoo} is the probability, that interval $S$ contains no energy level of a~simple sequence following GUE.

The evolution of level-spacing distributions, presented in Fig.\ \ref{psfi}, illustrates the fact, that for finite magnetic fields the effective Hamiltonian ${\cal H}_{\rm Dirac}$ (\ref{hameff}) does not commute with the symmetry ${\cal T}_v$ (\ref{tsltv}) and thus the valley-blocks are not degenerate. For finite systems, approximate valley degeneracy remains for $\Phi\ll{}\Phi_0$ (as the valley energy splitting is much smaller than the average level spacing). For $\Phi\gtrsim\Phi_0$, dynamical phases gained by carriers at $K$ and $K'$ valleys passing a~typical closed trajectory start to differ significantly \cite{Rec07foo}, and the sequences of energy levels corresponding to different valleys may be regarded as statistically-independent. 

It is worth to stress here, that the effect which we describe may appear for real magnetic fields only. In contrast, strain-induced pseudo-magnetic fields are exactly opposite at $K$ and $K'$ valleys, so they do not lift the valley degeneracy \cite{Voz10,Suz02}. We have found numerically, using the strained geometry of Ref.\ \cite{Low10}, that level-spacing distributions of chaotic nanosystems (each having the main symmetry axis bent into a piece of arc of the radius $R$) are unaffected even for extreme strains corresponding to $R/H=1$. The details of the calculations will be presented elsewhere.

\section{Spectral statistics in the localization range}
\label{wlocal}

\begin{figure}
\centerline{\includegraphics[width=\linewidth]{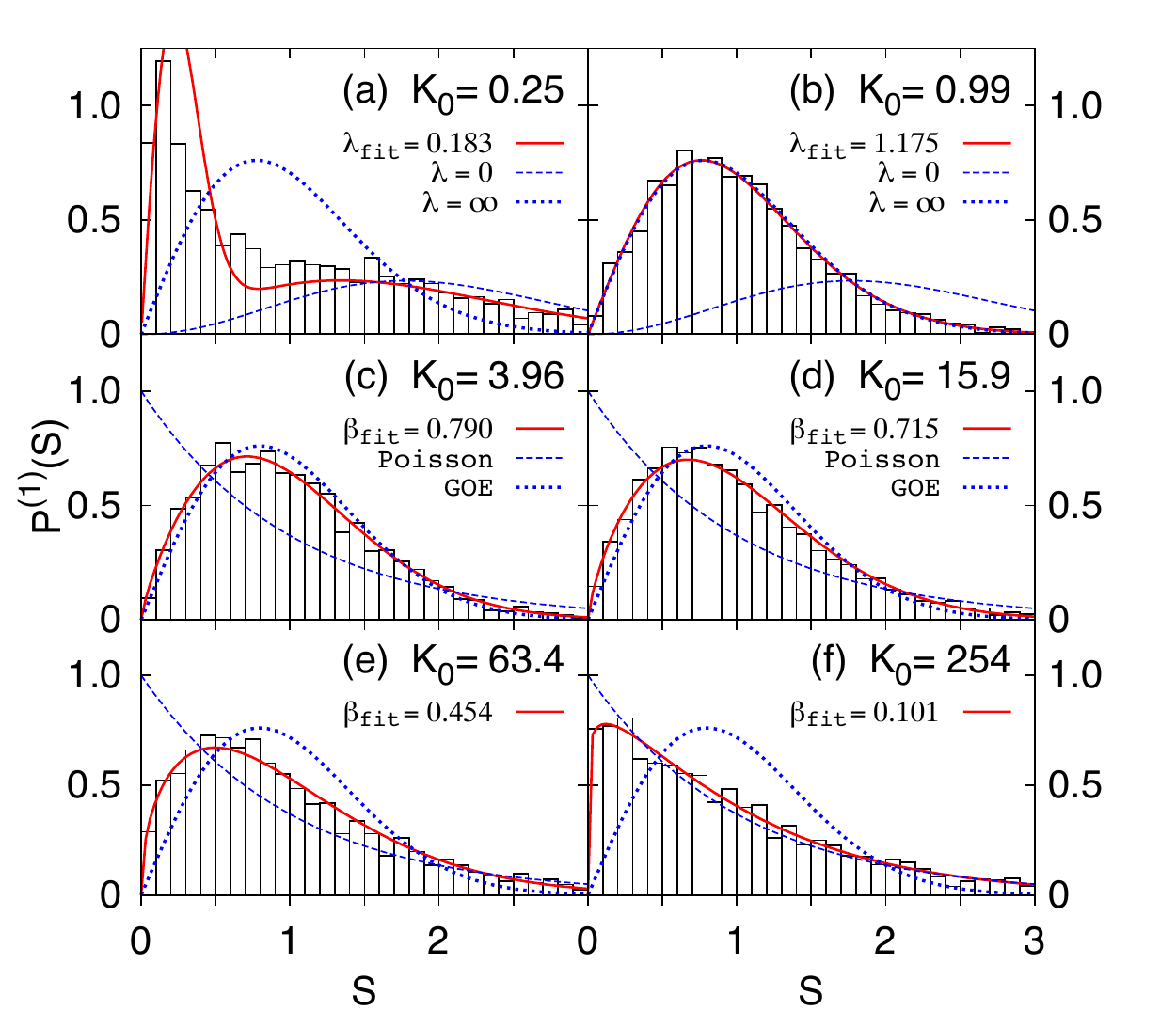}}
\caption{\label{pslarge} 
  Nearest-neighbor spacing distribution $P^{(1)}(S)$ for triangular nanoflakes with zigzag edges, the area ${\cal A}=(120\,a)^2$ \cite{lerefoo}, and large values of the disorder strength $K_0$ (specified at each panel). Numerical results are shown with black solid lines at all panels. The other lines at panel (a) and (b) are the same as in Figs.\ \ref{psnvac}(a)--(c). Red solid lines at panels (c)--(f) correspond to the Brody distribution $P_{\rm Brody}(\beta;S)$ (\ref{psbrody}) with least-square fitted $\beta=\beta_{\rm fit}$; blue lines are for the Poisson (dashed) and GOE (dotted) distributions.
}
\end{figure}

\begin{figure}
\centerline{\includegraphics[width=0.8\linewidth]{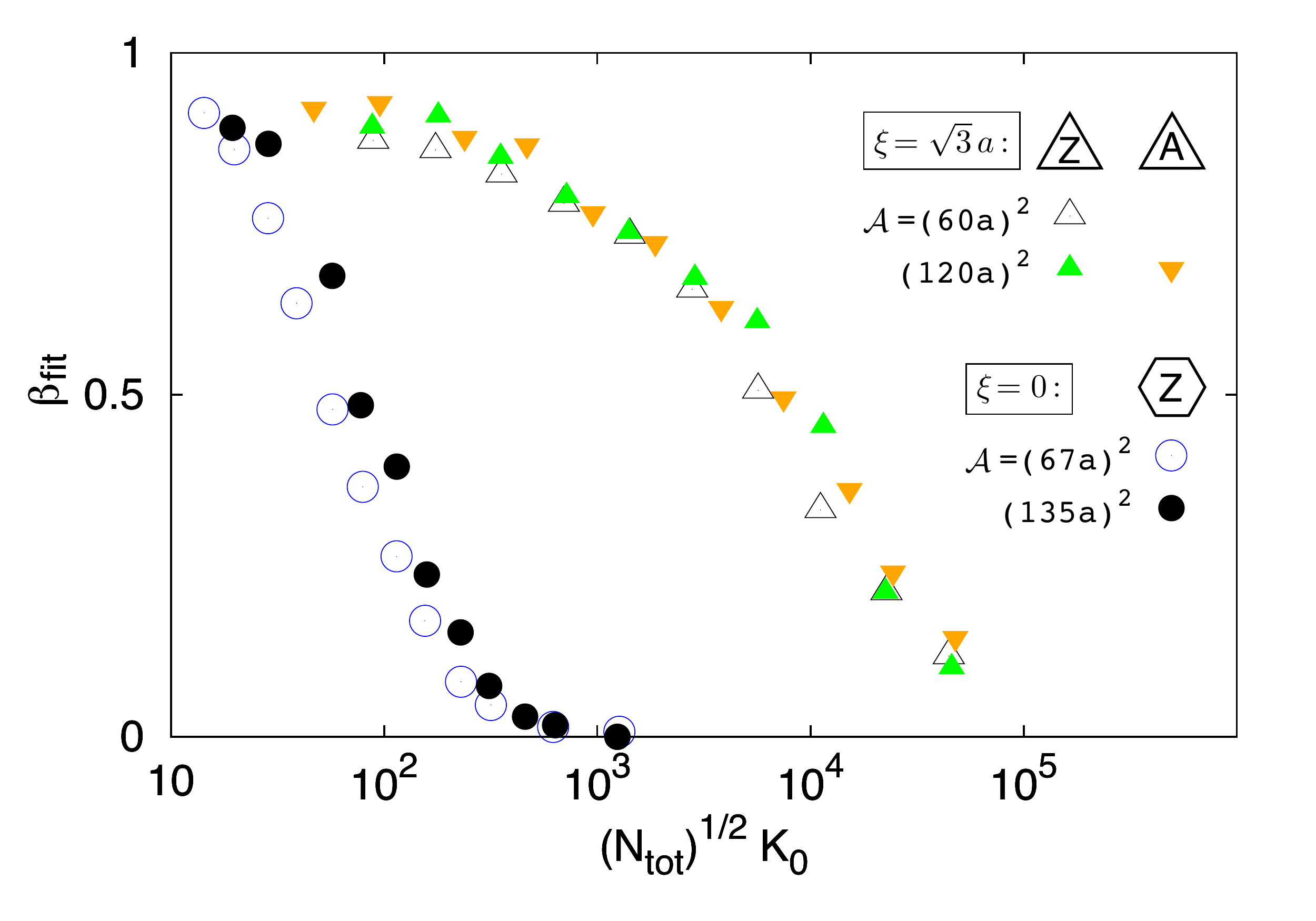}}
\caption{\label{betlarge} 
  Least-squares fitted parameters $\beta_{\rm fit}$ of Brody distributions $P_{\rm Brody}(\beta;S)$ (\ref{psbrody}) as functions of disorder strength for selected nanosystems of Table~\ref{sizetab} in the localization range. Datapoints correspond to triangles with zigzag ({\color{black}$\triangle$} $N_{\rm tot}=8278$, {\large\color{green}$\blacktriangle$} $N_{\rm tot}=32758$) or armchair ({\large\color{Orange}$\blacktriangledown$} $N_{\rm tot}=32760$) edges and smooth impurity potential ($\xi=\sqrt{3}\,a$). The results for hexagons with zigzag edges and Anderson-type disorder ($\xi=0$) are also shown ({\color{blue}$\bigcirc$} $N_{\rm tot}=10584$, {\Large$\bullet$} $N_{\rm tot}=42336$).
}
\end{figure}

So far, the issue of the wavefunction localization in chaotic nanosystems in graphene has been addressed numerically in the literature by employing models of disorder abruptly varying on the scale of atomic separation \cite{Xio07}. In this Section, we complement the existing studies with spectral statistics following from smooth impurity potential given by Eqs.\ (\ref{hamtba},\ref{uimper}). The scope of the paper is thus extended on the examples of highly-symmetric graphene nanoflakes with strong potential disorder.

It is know that in a generic quantum chaotic system eigenfunctions may not be uniformly distributed over a classically allowed phase space but localized (the dynamical localization effect), which is associated with the {\em fractional} power-law level repulsion \cite{Izr90}. In the presence of TRS, one can expect the crossover from GOE in the case when extended chaotic states dominate the spectrum to the Poisson distribution in the strong localization limit \cite{Bat10}. The level spacing distribution for the system undergoing such a~transition can be rationalized with the well-known Brody distribution 
\begin{equation}
\label{psbrody}
  P_{\rm Brody}(\beta;S)=C_1S^{\beta}\exp(-C_2S^{\beta+1}),
\end{equation}
with 
\begin{equation}
  C_1=(\beta+1)C_2,\ \ \ \ 
  C_2=\left[\Gamma\left(\frac{\beta+2}{\beta+1}\right)\right]^{\beta+1},
\end{equation}
and $\Gamma(x)$ being the Gamma function. The limiting distributions $P_{\rm GOE}(S)$ (\ref{pgoe}) or $P_{\rm Poi}(S)=\exp(-S)$ are restored for $\beta=1$ or $\beta=0$ (respectively). In turn, when analyzing the spectral statistics of strongly disordered and closed nanosystem in graphene that preserves TRS, one can fit the empirical distribution $P_{\rm Brody}(\beta;S)$ (\ref{psbrody}) to the actual nearest-neighbor spacing distribution $P^{(1)}(S)$ in order to quantify the deviations from $P_{\rm GOE}(S)$, which indicate the localization. 

At sufficiently strong disorder all the systems studied in the paper show intervalley scattering which restores TRS (at zero magnetic field). In particular, the spacing distributions $P^{(1)}(S)$ obtained numerically for triangular nanoflakes with zigzag edges ($N_{\rm vac}=0$), the area ${\cal A}=(120\,a)^2$, and different values of the disorder strength $K_0$ \cite{lerefoo} are depicted with black solid lines in Fig.\ \ref{pslarge}. When increasing $K_0$, the distribution $P^{(1)}(S)$ first show a crossover from the theoretical prediction for GUE with twofold valley degeneracy [blue dashed lines in Figs.\ \ref{pslarge}(a) and \ref{pslarge}(b)] given by Eq.\ (\ref{psak1zero}) to the Wigner surmise for GOE [blue dotted lines], approaching the latter at $K_0\simeq{}1$. We notice here, that for the system parameters given in Ref.\ \cite{lerefoo} $K_0\simeq{}1$ corresponds to the disorder amplitude $\delta/t\simeq{}0.5$, i.e., the orthogonal symmetry manifests itself in $P^{(1)}(S)$ if $|E|\lesssim{}\delta$ for all energy levels from the range $0.1\leqslant{}|E|/t\leqslant{}0.5$, which are taken into account. Interestingly, in the crossover range ($K_0\lesssim{}1$) $P^{(1)}(S)$ can still be rationalized with the empirical distribution $P_{\overline\alpha,\overline\kappa}^{(1)}(\lambda;S)$ (\ref{ps1barak}) with least-square fitted $\lambda=\lambda_{\rm fit}$ [red solid lines in Figs.\ \ref{pslarge}(a) and \ref{pslarge}(b)], similarly as in the case of transition GUE-GOE observed when increasing the number of edge vacancies $N_{\rm vac}$ (see Sec.\ \ref{tragogu}). For $K_0\gg{}1$, the distribution $P^{(1)}(S)$ can be approximated by the Brody ditribution $P_{\rm Brody}(\beta;S)$ (\ref{psbrody}) with least-squares fitted $\beta=\beta_{\rm fit}$ [red solid lines in Figs.\ \ref{pslarge}(c)--\ref{pslarge}(f)] and gradually approaches the Poisson distribution [blue dashed lines] signalling that the chaotic but localized eigenstates start to govern the spectrum.

The evolution of $P^{(1)}(S)$ illustrated in Figs.\ \ref{pslarge}(c)--\ref{pslarge}(f) is qualitatively reproduced for all nanosystems considered (see Table~\ref{sizetab}) in their localization ranges. The value $K_0$, at which the crossover from GOE to the Poisson distribution occurs, is however related to the system size and microscopic details of the disorder. For a quantitative description we plot, in Fig.\ \ref{betlarge}, the values of $\beta_{\rm fit}$ in the Brody distribution (\ref{psbrody}) best-fitted to the actual distributions $P^{(1)}(S)$ obtained numerically for different nanosystems, as functions of the dimensionless quantity $N_{\rm tot}^{1/2}K_0$. After such a scaling of the independent variable, the available datasets group around just two distinct curves, one for the Anderson-type disorder characterized by $\xi=0$ (circles in Fig.\ \ref{betlarge}), and the other for smooth impurity potential with $\xi=\sqrt{3}\,a$ (remaining symbols). We attribute it to the fact, that $N_{\rm tot}^{1/2}K_0$ is proportional to the effective system size $\sqrt{\cal A}/\langle{}l(E)\rangle$ determined by the free path $l(E)$ in Born approximation \cite{Sch98}
\begin{equation}
  \label{leborn}
  l(E)=\frac{2g_v\hbar{}v_F}{K_0|E|}, \ \ \ \ 
  g_v = \begin{cases}
    1, & \text{if } \xi\ll{}a, \\
    2, & \text{if } \xi\gtrsim{}a.
  \end{cases}
\end{equation}
Averaging $l(E)$ over the range $E_{\rm min}\leqslant|E|\leqslant{}E_{\rm max}$, with the weights $\rho_{\rm bulk}(E)$ given by Eq.\ (\ref{rhobulk}), one gets
\begin{equation}
  N_{\rm tot}^{1/2}K_0=
  \frac{2\cdot{}3^{1/4}g_vt}{E_{\rm min}\!+\!E_{\rm max}}
  \,\frac{\sqrt{\cal A}}{\langle{}l(E)\rangle}
  \simeq{}4.4\,g_v\frac{\sqrt{\cal A}}{\langle{}l(E)\rangle},
\end{equation}
where the last approximate equality holds true for $E_{\rm min}=0.1\,t$ and $E_{\rm max}=0.5\,t$ used in our numerical simulations. 

It is also visible from Fig.\ \ref{betlarge}, that the crossover to the localization range takes place for in case of $\xi=\sqrt{3}\,a$ for $N_{\rm tot}^{1/2}K_0$ more than two orders of magnitude larger than in the case of $\xi=0$. This suggests, that the smooth character of the disorder potential may be crucial for experimental observation of signatures of quantum chaos in closed graphene nanosystems.

\section{Conclusions \label{conclu}}
We have investigated the symmetry classes of selected closed nanosystems in graphene (equilateral triangles and hexagons with three types of boundaries: armchair, zigzag, and Klein) and studied the effect of weak potential disorder. Predictions of the Dirac equation for low-energy excitations were confronted with numerical results for the tight-binding model on a~honeycomb lattice. New findings are visualized in Fig.\ \ref{gradyn}.

In the absence of disorder, available analytic solutions for continuous Dirac billiards show the level clustering due to number-theoretic degeneracies. Such degeneracies are lifted up in the tight-binding model due to nonlinear terms in the dispersion relation, leading to the Poissonian distribution of energy levels (characteristic for a generic integrable system).

For weak disorders, the transition to quantum chaos is observed. In such a limit, spectral statistics follow these characterizing Gaussian ensembles of random matrices. In principle, all of the considered tight-binding Hamiltonians are time-reversal invariant and expected to show the orthogonal symmetry class in the absence of magnetic field. To the contrary, the Dirac Hamiltonian for graphene has a block structure related to the presence of two valleys making the true TRS irrelevant in the absence of intervalley scattering. Instead, special TRS (symplectic symmetry) applies and is broken by the disorder, leading to the unitary class (accompanied by the twofold valley degeneracy of each level). In effect, the type of boundaries plays a decisive role for the symmetry properties.

Earlier studies of closed nanosystems in graphene \cite{Wur09,Der08,Lib09,Ryc11} reported the orthogonal symmetry class associated with the valley mixing. We have found that the unitary symmetry class can also be observed in spectral statistics of such systems, providing almost all terminal atoms belong to one sublattice. This is satisfied for equilateral triangles with zigzag or Klein boundaries, for which the spectral statistics obtained numerically show the following features: When increasing the disorder strength, transition from the Poisson to GUE distribution (both showing an approximate twofold degeneracy of each level) occurs. For a fixed disorder strength in the chaotic range and increasing the number of edge vacancies we have observed the transition to GOE distribution (accompanied by the gradual level splitting). Moreover, for the same disorder strength and in the absence of edge vacancies, we have demonstrated the evolution from GUE distribution with twofold degeneracies to the distribution characterizing two independent GUEs at weak magnetic fields. These findings complement the very recent results \cite{Wur11b} for transport characteristics of open nanosystems in graphene.

The remaining nanosystems studied in the paper are in the orthogonal symmetry class. For all cases, the transition to quantum chaos is rationalized using additive random-matrix models.  The functional relation between the best-fitted model parameter $\lambda_{\rm fit}$ and the disorder strength has a form of a power law $\lambda_{\rm fit}\propto{}(N_{\rm tot}K_0)^\alpha$, with the symmetry-dependent exponent $\alpha$ taking different values for systems undergoing the transitions Poisson-GOE and Poisson-GUE. Additionally, the model involving $4\times{}4$ real random matrices is proposed and elaborated to parametrize the nonstandard GUE-GOE transition identified for triangular flakes with edge vacancies at zero magnetic field. For strong disorders (i.e., in the localization range) additive random-matrix models no longer apply. Instead, the fractional level repulsion and the evolution, with the increasing disorder strength, towards the Poissonian distribution of energy levels are observed, indicating the spacial localization of quantum states.

We hope recent progress in resolving closely-lying energy levels in graphene quantum dots using the three-terminal Coulomb-blockade setup \cite{Jac12} will make it possible to test experimentally our results.

\section*{Acknowledgments}
I thank to Grzegorz Rut for remarks on a~role of the Zeeman effect in graphene. Discussions with \.{I}nan\c{c} Adagideli, Klaus Richter, J\"{u}rgen Wurm, and Karol \.{Z}yczkowski are appreciated. The work was supported by the National Science Centre of Poland (NCN) via Grant No.\ N--N202--031440, by the Alexander von Humboldt Foundation (AvH), and partly by Foundation for Polish Science (FNP) under the program TEAM.


\appendix

\section{Boundary conditions to Dirac equation and symmetry classes of graphene nanoflakes \label{bocodi}}

Boundary conditions for Dirac fermions in graphene are usually discussed in the so-called {\em valley-isotropic representation} \cite{Akh07}. In this Appendix, we recall the standard expressions  \cite{Bee08,Akh08} for infinite mass, armchair, zigzag and Klein boundaries, and rewrite them in the notation of Eq.\ (\ref{hameff}) to illustrate how particular boundaries may determine the system symmetry class. 

The valley-isotropic representation of the Hamiltonian (\ref{hameff}) can be obtained using the following unitary transformation \cite{Akh08,Wur11a}
\begin{multline}
  \label{tilham}
  \tilde{\cal H}_{\rm Dirac}={\cal U}^{\dagger}{\cal H}_{\rm Dirac}\,{\cal U}=
  v_F\left[\left(\mbox{\boldmath$p$}+e\mbox{\boldmath$A$}\right)\cdot
    \mbox{\boldmath$\sigma$}\right]\otimes\tau_0 \\ + 
  M({\bf r})\sigma_z\otimes\tau_z + U({\bf r})\sigma_0\otimes\tau_0,
\end{multline}
where
\begin{multline}
  \label{calu}
  {\cal U}=\left(\begin{array}{cccc}
      1 & 0 & 0 & 0 \\
      0 & 1 & 0 & 0 \\
      0 & 0 & 0 & 1 \\
      0 & 0 & -1 & 0 \\
    \end{array}\right) = \\
  \frac{1}{2}\sigma_0\otimes(\tau_0+\tau_z)
  + \frac{i}{2}\sigma_y\otimes(\tau_0-\tau_z),
\end{multline}
and the remaining symbols are the same as in Eq.\ (\ref{hameff}). The Hamiltonian (\ref{tilham}) now acts on spinors  $\tilde{\psi}\equiv{\cal U}^{\dagger}\psi=[\psi_A,\psi_B,-\psi_B',\psi_A']^T$. For model situations considered in the literature \cite{Akh08,Wur11a} the mass term $M({\bf r})=0$ and the Hamiltonian (\ref{tilham}) contains only terms proportional to $\tau_0$, so it consists of two identical blocks (one for each valley) justifying the notion of 'valley-isotropic representation'. As we show in Section \ref{dismod}, the potential disorder in graphene leads to $M({\bf r})\neq{0}$, so the term proportional to $\tau_z$ appears in the Hamiltonian. However, the representation (\ref{tilham}--\ref{calu}) still remains useful for defining the boundary conditions to Dirac equation. Also, the current operator 
\begin{equation}
  \label{tilj}
  \tilde{\mbox{\boldmath$\jmath$}}=
  v_F\mbox{\boldmath$\sigma$}\otimes\tau_0
\end{equation}
is proportional to $\mbox{\boldmath$\sigma$}\otimes\tau_0$ and thus has identical form for both valleys, regardless $M({\bf r})=0$ or $M({\bf r})\neq{0}$ \cite{natj}.

Most common boundary conditions for graphene nanosystems may be expressed in a compact form as \cite{Akh07}
\begin{equation}
  \label{tilmbc}
  \tilde{\psi}=\tilde{\cal M}\tilde{\psi}, \ \ \ 
  \tilde{\cal M} = 
  \left(\mbox{\boldmath$\eta$}\cdot\mbox{\boldmath$\sigma$}\right) \otimes 
      \left(\mbox{\boldmath$\nu$}\cdot\mbox{\boldmath$\tau$}\right),
\end{equation}
where $\mbox{\boldmath$\eta$}$ and $\mbox{\boldmath$\nu$}$ are three-dimensional unit vectors. The vector  $\mbox{\boldmath$\eta$}$ is constrained by $\mbox{\boldmath$\eta$}\cdot\mbox{\boldmath$\eta$}_B$ to guarantee that no current leaks out of the boundary, defined by the normal $\mbox{\boldmath$\eta$}_B$ (pointing outward). In fact, Eq.\ (\ref{tilmbc}) represents the general boundary condition, providing we ignore noncollinear local magnetization (which may appear on the edges of a graphene nanoflake \cite{Sep10,Voz11}), so one can assume that the boundary condition itself does not break time-reversal symmetry \cite{Bee08}.

The examples of boundaries that can be defined via vectors  $\mbox{\boldmath$\eta$}$ and $\mbox{\boldmath$\nu$}$ in Eq. (\ref{tilmbc}) are:
\begin{itemize}
\item
Confinement by an infinite mass corresponds to $\mbox{\boldmath$\eta$}=\pm\,\hat{\bf e}_z\times\mbox{\boldmath$\eta$}_B$ and $\mbox{\boldmath$\nu$}=\hat{\bf e}_z$, where the upper (lower) sign is valid for the mass going to $+\infty$ ($-\infty$) outside the system area.
\item
An armchair edge requires the wavefunction (\ref{wavefun}) is vanishing on both sublattices, namely: $\psi_X\exp\left(i{\bf K}\cdot{\bf r}\right)+\psi'_X\exp\left(-i{\bf K}\cdot{\bf r}\right)=0$ for $X=A,B$. This corresponds to $\mbox{\boldmath$\eta$}=\pm\,\hat{\bf e}_z\times\mbox{\boldmath$\eta$}_B$ and $\hat{\bf e}_z\cdot\mbox{\boldmath$\nu$}=0$, where the upper (lower) sign is valid when the order of the atoms within each dimer is $A\!-\!B$ ($B\!-\!A$) along the direction of $\hat{\bf e}_z\times\mbox{\boldmath$\eta$}_B$.
\item
A zigzag (or Klein) edge requires $\psi_A=\psi_A'=0$ or $\psi_B=\psi_B'=0$, depending on whether the row of missing atoms at the edge is on the $A$ or $B$ sublattice, what corresponds to $\mbox{\boldmath$\eta$}=\pm\,\hat{\bf e}_z$ and $\mbox{\boldmath$\nu$}=\hat{\bf e}_z$.
\end{itemize}

The boundary condition (\ref{tilmbc}) can be written in the notation of Eq.\ (\ref{hameff}) as 
\begin{equation}
  \label{natpsibc}
  \psi=\left({\cal U}\tilde{\cal M}{\cal U}^{\dagger}\right)\psi
  \equiv{\cal M}\psi,
\end{equation}
where 
\begin{multline}
  {\cal M}= 
  \left(\eta_x\sigma_x+\eta_y\sigma_y\right)\otimes\nu_z\tau_0  
  + \eta_y\sigma_y\otimes\nu_z\tau_z \\ 
  + \left(\eta_x\sigma_z-n_z\sigma_x\right)\otimes
  \left(\nu_x\tau_x+\nu_y\tau_y\right) \\
  + \eta_y\sigma_0\otimes\left(\nu_x\tau_y-\nu_y\tau_x\right).
\end{multline}
For the most common boundary conditions listed above, ${\cal M}$ is given by
\begin{multline}
  \label{natmbc}
  {\cal M}= \\
  \begin{cases}
      \eta_x\sigma_x\otimes\tau_0 + \eta_y\sigma_y\otimes\tau_z
      & \text{(infinite mass)} \\
      \eta_x\sigma_z\otimes\left(\nu_x\tau_x\!+\!\nu_y\tau_y\right) \\
      \ \ +\,\eta_y\sigma_0\otimes\left(\nu_x\tau_y\!-\!\nu_y\tau_x\right)
      & \text{(armchair)} \\
      \pm\,\sigma_z\otimes\tau_0 & \text{(zigzag/Klein)}
  \end{cases},
\end{multline}
where $\mbox{\boldmath$\eta$}=(\eta_x,\eta_y)=\pm\,\hat{\bf e}_z\times\mbox{\boldmath$\eta$}_B$ for the first two cases. For armchair edge, $\mbox{\boldmath$\nu$}=(\nu_x,\nu_y)$ is a unit vector in the $x\!-\!y$ plane.

\begin{figure}
\centerline{\includegraphics[width=0.8\linewidth]{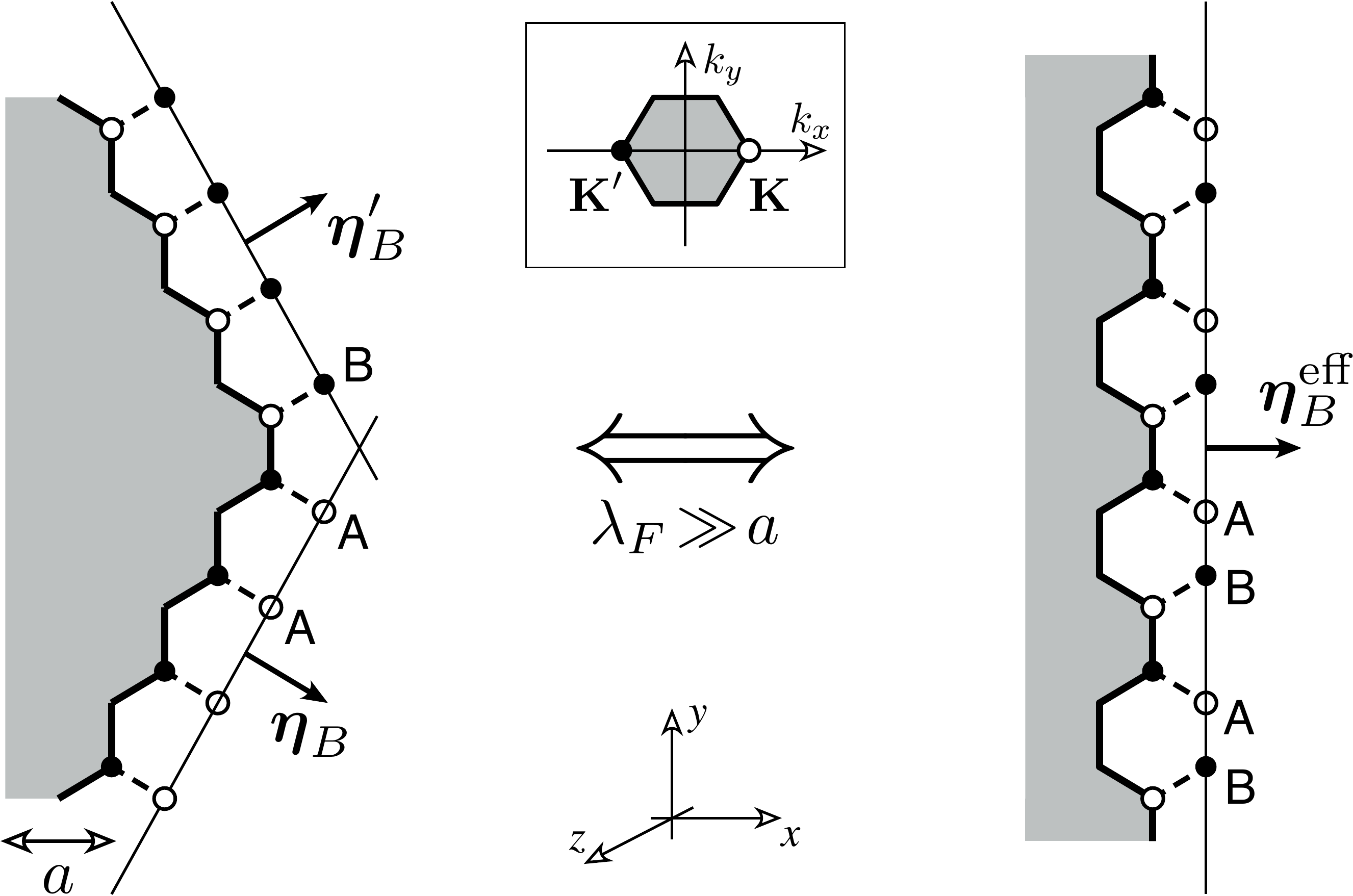}}
\caption{\label{zigvfig}
  {\em Left}---two zigzag edges of different types forming a $120^\circ$ corner, {\em right}---an armchair edge (tick solid lines). These two boundaries become equivalent if an electron (or hole) incoming from the sample area (shaded) has a~Fermi wavelength $\lambda_F$ large compares to the lattice spacing $a$. The atoms in terminal and first missing rows are indicated with open or closed dots depending if they belong to $A$ or $B$ sublattice. (Each dashed line marks a~bond between the terminal and the nearest missing atom.) $\mbox{\boldmath$\eta$}_B$, $\mbox{\boldmath$\eta$}_B'$, and $\mbox{\boldmath$\eta$}_B^{\rm eff}$ are unit vectors normal to the boundaries. The coordinate system and the first Brillouin zone (inset) are also shown.  
} 
\end{figure}

It is clear from Eq.\ (\ref{natmbc}) that the boundary condition (\ref{natpsibc}) couples the valley degree of freedom for the case of armchair edges, leading to the orthogonal symmetry class of a~chaotic nanosystem at zero magnetic field. For the remaining cases, no obvious intervalley scattering originates from the edges, so one may expect that the unitary symmetry class appears. The more carefully discussion is necessary, however, for systems containing two zigzag (or Klein) edges of different types. 

In particular, we focus here on $120^\circ$ corners formed by a~zigzag edge terminated on $A$ sublattice attached to a~zigzag edge terminated on $B$ sublattice (such as presented in Fig.\ \ref{zigvfig}), which appear in hexagons with zigzag edges. An effective wavefunction for low-energy excitations (\ref{wavefun}) near the lower arm of such a~corner is determined by the boundary condition that may be written as $\psi_A\exp\left(i{\bf K}\cdot{\bf r}\right)+\psi'_A\exp\left(-i{\bf K}\cdot{\bf r}\right)=0$, where ${\bf r}$ denotes the position at the first row of missing atoms. This condition cannot be satisfied simply by setting $\psi_A=\psi_A'=0$, because the wavefunction also needs to satisfy an analogous condition for the upper arm, namely: $\psi_B\exp\left(i{\bf K}\cdot{\bf r}'\right)+\psi'_B\exp\left(-i{\bf K}\cdot{\bf r}'\right)=0$. In the coordinate system of Fig.\ \ref{zigvfig}, we immediately get that these two conditions lead to
\begin{gather}
 \mbox{\boldmath$\eta$}=\hat{\bf e}_z\times\mbox{\boldmath$\eta$}_B^{\rm eff},
 \ \ \ 
 \hat{\bf e}_z\cdot\mbox{\boldmath$\nu$}=0, \nonumber \\
 \mbox{\boldmath$\eta$}_B^{\rm eff}\equiv
 \frac{ \mbox{\boldmath$\eta$}_B+\mbox{\boldmath$\eta$}_B' }{%
   \|\mbox{\boldmath$\eta$}_B+\mbox{\boldmath$\eta$}_B'\| }.
\end{gather}
This is an effective armchair boundary condition. 

We can expect now, that strong intervalley scattering originating from $120^\circ$ corners places hexagons with zigzag edges in the orthogonal symmetry class. Similar arguments apply to hexagons with Klein edges. For this reason, the only closed systems for which the unitary symmetry class may still manifests itself in spectral statistics are those bounded entirely with zigzag (or Klein) edges with terminal atoms belonging to one sublattice, such as equilateral triangles considered in Sections \ref{leretri}--\ref{tragogu}.

\section{Low-energy level structure of triangular graphene flakes
  \label{loentri}}

\begin{table}
\caption{\label{emnatritab}
  Lowest-lying electronic levels ($E_{m,n,+}^{\rm Dirac}$) obtained from Eq.\ (\ref{emndiractri}) for triangular graphene flake with armchair edges containing $N_{\rm tot}=32760$ atoms and corresponding energies ($E$) obtained from numerical diagonalization of ${\cal H}_{\rm TBA}$ (\ref{hamtba}) for zero and for weak disorder (specified by $K_0$). The staggered potential $M_V({\bf r}_i)=0$ for all lattice sites.}
\begin{tabular}{c|c|c|c}
  \hline\hline
  $(m,n)$ & $E_{m,n,+}^{\rm Dirac}/t$ & $E/t$, & $E/t$, \\ 
  &  & $K_0\!=\!0$ &  $K_0\!=\!8\!\times\!10^{-4}$ \\ \hline  
  &  &  &  \\
  (1,1) &  $\ $0.02004229$\ $ & $\ $0.01994629$\ $ & 0.01992564 \\ 
  &  &                  0.01994629 & 0.01992624 \\
  &  &  &  \\
  (1,2) &  0.03471426 & 0.03434734 & 0.03410399 \\
  &  &                  0.03474520 & 0.03450004 \\   
  &  &  &  \\
  (2,2) &  0.04008458 & 0.03989059 & 0.03959939 \\
  &  &                  0.03989059 & 0.03960217 \\
  &  &  &  \\
  (2,3) &  0.05302691 & 0.05237556 & 0.05185480 \\
  &  &                  0.05315689 & 0.05271721 \\
  (1,3) &  0.05302691 & 0.05237556 & 0.05225943 \\
  & &		        0.05315689 & 0.05312196 \\
  &  &  &  \\
  (3,3) &  0.06012686 &	0.05983093 & 0.05990241 \\
  &  &	                0.05983093 & 0.05990582 \\
  &  &  &  \\
  (2,4) &  0.06942852 &	0.06828316 & 0.06797621 \\
  &  &   	 	0.06987442 & 0.06954428 \\
  &  &  &  \\
  (3,4) &  0.07226350 &	0.07132749 & 0.07112046 \\
  &  &     	  	0.07247406 & 0.07228640 \\
  (1,4) &  0.07226350 & 0.07132749 & 0.07156919 \\
  &  & 		        0.07247406 & 0.07272521 \\
  &  &  &  \\
  (4,4) &  0.08016915 &	0.07976532 & 0.07971228 \\
  &  &  		0.07976532 & 0.07971380 \\
  &  &  &  \\
  (2,5) &  0.08736231 & 0.08572437 & 0.08523396 \\
  &  &    		0.08809530 & 0.08767157 \\
  (3,5) &  0.08736231 &	0.08572437 & 0.08566093 \\
  &  &	  		0.08809530 & 0.08811008 \\
  &  &  &  \\
  (4,5) &  0.09184530 & 0.09062032 & 0.09035237 \\
  &  & 	  	      	0.09212365 & 0.09184820 \\
  (1,5) &  0.09184530 & 0.09062032 & 0.09059671 \\
  &  & 	  		0.09212365 & 0.09209424 \\
  &  &  &  \\
  (5,5) &  0.10021144 &	0.09969177 & 0.09933460 \\
  &  & 	  		0.09969177 & 0.09936139 \\
  &  &  &  \\ \hline\hline
\end{tabular}
\end{table}

\begin{table}
\caption{\label{emntrigtab}
  Same as Table \ref{emnatritab}, but for triangular flake with zigzag edges and $N_{\rm tot}=32758$. }
\begin{tabular}{c|c|c|c}
\hline\hline
  $(m,n)$ & $E_{m,n,+}^{\rm Dirac}/t$ & $E/t$, & $E/t$, \\ 
  &  & $K_0\!=\!0$ &  $K_0\!=\!6\!\times\!10^{-4}$ \\ \hline  
  &  &  &  \\
  (1,2) &  $\ $0.03471532$\ $ & $\ $0.03452128$\ $ & 0.03437910 \\
  &  &  	        0.03452128 & 0.03437919 \\
  &  &  &  \\
  (1,3) &  0.05302853 & 0.05247781 & 0.05208748 \\
  &  & 	  		0.05247781 & 0.05208754 \\
  &  &			0.05297828 & 0.05301960 \\
  &  &		        0.05297828 & 0.05301991 \\
  &  &  &  \\
  (2,4) &  0.06943064 &	0.06903228 & 0.06873980 \\
  &  & 	  		0.06903228 & 0.06874017 \\
  &  &  &  \\
  (1,4) &  0.07226570 & 0.07120437 & 0.07120256 \\
  &  & 	  		0.07120437 & 0.07120295 \\
  &  &		        0.07248965 & 0.07228947 \\
  &  &			0.07248965 & 0.07228985 \\
  &  &  &  \\
  (2,5) &  0.08736497 & 0.08642740 & 0.08616840 \\
  &  & 	  		0.08642740 & 0.08616900 \\
  &  &	                0.08727781 & 0.08701618 \\
  &  &			0.08727781 & 0.08701620 \\
  &  &  &  \\
  (1,5) &  0.09184810 & 0.09012997 &   0.09000825 \\
  &  & 	  		0.09012997 &   0.09000841 \\
  &  &	       	        0.09247007 &   0.09214693 \\
  &  &		        0.09247007 &   0.09214700 \\
  &  &  &  \\ \hline\hline
\end{tabular}
\end{table}

In Tables \ref{emnatritab} and \ref{emntrigtab} we list energy levels $E_{m,n,+}^{\rm Dirac}$ (\ref{emndiractri}) of two triangular nanoflakes considered in the paper (see Table~\ref{sizetab}), by taking $n\leqslant{}5$ and $m$-s satisfying (\ref{mnatri}) or (\ref{mntrig}) for armchair or zigzag edges (respectively). Corresponding energies obtained from the exact numerical diagonalization of tight-binding Hamiltonians (\ref{hamtba}) in the absence of disorder ($K_0=0$) and for infinitesimally-weak bulk disorder ($K_0\simeq{}10^{-3}$; for the disorder details see Ref.\ \cite{csklfoo}) are also provided. Unlike for similar Schr\"{o}dinger systems \cite{Roz10,Kri82}, the spinor structure of the wavefunction (\ref{wavefun}) cause that geometric symmetries of graphene flakes do not lead directly to level degeneracies. Instead, degeneracies associated with special time-reversal symmetries ${\cal T}_{\rm sl}$ and ${\cal T}_v$ (\ref{tsltv}) may appear.

We see from Table~\ref{emnatritab}, that for armchair boundaries each electronic level of the Dirac cavity $E_{m,n,+}^{\rm Dirac}$ is followed by two levels of the lattice system, corresponding to the Kramer's degeneracy associated with ${\cal T}_{\rm sl}$ (\ref{tsltv}). For the lattice system, this degeneracy is usually only approximate even at $K_0=0$, as a~nonlinear term appearing in the effective Hamiltonian derived from tight-binding model does not commute with ${\cal T}_{\rm sl}$  \cite{And07}. Analogous symmetry ${\cal T}_v$ (\ref{tsltv}) does not apply for armchair edges, however, the property $E_{m,n}^{\rm Dirac}=E_{n-m,n}^{\rm Dirac}$ leads to an additional twofold degeneracy of each level, providing that $m\neq{}n$ and $2m\neq{}n$. As a~result, for $K_0=0$ majority of energy levels occurs in almost-degenerated quadruplets. The degeneracy is immediately lifted in the presence of disorder. Typically, as small disorder strength as $K_0\simeq{}10^{-3}$ leads to the level splitting $\delta{}E\gtrsim{}10^{-4}\,t$ for $|E|\lesssim{}0.1\,t$. 

For zigzag edges and for $M_V\equiv{}0$ (see Table~\ref{emntrigtab}), the fourfold (approximate for the finite lattice system) degeneracy appears for almost every level, due to the symmetries ${\cal T}_{\rm sl}$ and ${\cal T}_v$ (\ref{tsltv}). For even $n$, the degeneracy is only twofold when $2m=n$. We attribute this to the presence of edges states in the system with zigzag edges, which have missing valley degeneracies of bulk states. Such levels, however, do not contribute to spectral statistics of large systems. Unlike for armchair boundaries, the twofold valley degeneracy associated with the symmetry ${\cal T}_v$ (\ref{tsltv}) is present for almost every level of the triangle with zigzag edges, and appears to be very robust against the disorder. For  $K_0\simeq{}10^{-3}$, corresponding splittings from Table~\ref{emntrigtab} are $\delta{E}\lesssim{}10^{-7}\,t$ for $|E|\lesssim{}0.1\,t$. (Notice that the degeneracy associated with ${\cal T}_{\rm sl}$ is lifted in the presence of disorder for either zigzag or armchair boundary conditions.) These splittings are unaffected when the staggered potential with $|M_V({\bf r}_i)|=0.7\,t$ is put on the outermost edge atoms. The identical structure of energy levels as for zigzag edges was observed for the case of a triangle with Klein edges. 

These are the reasons, why large and weakly-disordered triangular nanoflake in graphene shows the twofold, approximate level degeneracy only if it has perfect zigzag (or Klein) edges. In other cases, no degeneracies appear in the presence of disorder.

\section{Level clustering in triangular graphene flakes
  \label{leclutri}}

Electronic energies of triangular Dirac cavities (\ref{emndiractri}) may be written, in the dimensionless units, as
\begin{multline}
\label{emndiracdim}
  \frac{E_{m,n,+}^{\rm Dirac}}{\Delta}=2\sqrt{m^2+n^2-mn}=\\
  \sqrt{(m+n)^2+3(m-n)^2}= \sqrt{l^2+3k^2}\equiv \epsilon_{kl},
\end{multline}
where $\Delta\equiv{}\pi{}t/\sqrt{3N_{\rm tot}}$, $k\equiv{}n-m$, and $l\equiv{}n+m$ \cite{kldefoo}. Without the loss of generality, we have limited the discussion to energy levels in the conduction band. Let $\epsilon_{kl}\geqslant\epsilon_{k'l'}$ are the nearest neighbors in the sequence. For high energies, i.e., $\epsilon_{kl},\epsilon_{k'l'}\gg\Delta$, we have
\begin{multline}
\label{eps2kl}
  \epsilon_{kl}^2-\epsilon_{k'l'}^2=l^2+3k^2-(l')^2-3(k')^2\simeq \\
  2\epsilon_{kl}\left(\epsilon_{kl}-\epsilon_{k'l'}\right)=
  \frac{6\sqrt{3}}{\pi}\tilde{\rho}(\epsilon_{kl})\left(\epsilon_{kl}-\epsilon_{k'l'}\right),
\end{multline}
where the dimensionless density of states $\tilde{\rho}(\epsilon)$ is defined via  $\rho_{\rm bulk}(E)\equiv{\Delta}^{-1}\tilde{\rho}(E/\Delta)$, with $\rho_{\rm bulk}(E)$ given by Eq.\ (\ref{rhobulk}). By definition, $\tilde{\rho}(\epsilon_{kl})\left(\epsilon_{kl}-\epsilon_{k'l'}\right)=S$. Denoting the integer in the first line of Eq.\ (\ref{eps2kl}) by
\begin{equation}
  q\equiv{} l^2+3k^2-(l')^2-3(k')^2
\end{equation}
we immediately obtain the quantization rule for $S$ given by Eq.\ (\ref{squant}) in the main text.

Similarly as for the analogous Schr\"{o}dinger systems \cite{Ber81,Pin80,Kri82}, energy levels of equilateral triangles containing Dirac fermions show number-theoretic degeneracies not connected to the geometric symmetry. In particular, the result by Pinsky \cite{Pin80} who showed that the average level multiplicity is divergent when expanding the energy interval from which the levels are taken into account applies directly to energy levels given by Eq.\ (\ref{emndiractri}). The divergence (or level clustering) also appears when fixing the energy range, i.e., $|E_{m,n,\pm}^{\rm Dirac}|\leqslant{}E_{\rm max}$, and increasing $N_{\rm tot}$. For these reasons, equilateral triangles with Dirac fermions cannot be regarded as generic integrable systems, unless nonlinear terms in the dispersion relation (originating from the tight-binding Hamiltonian of graphene) lift up number-theoretic degeneracies.

\begin{figure}
\centerline{\includegraphics[width=0.8\linewidth]{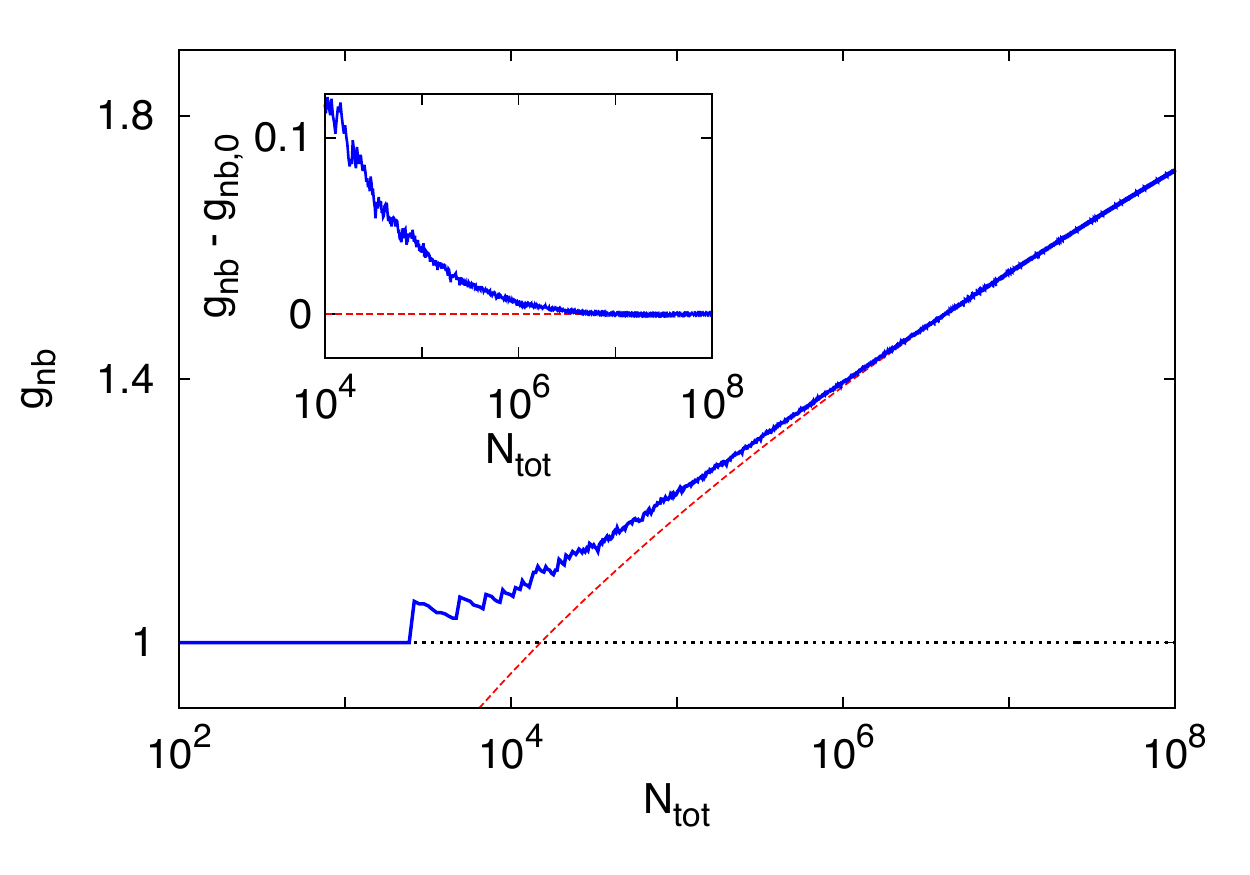}}
\caption{ \label{ndegatri}
  Average level degeneracy for a~triangular nanoflake with armchair edges containing $N_{\rm tot}$ atoms for energies $E_{m,n,\pm}^{\rm Dirac}$ obtained from Eq.\ (\ref{emndiractri}) [blue solid line], and $E_{m,n,\pm}^{\rm TBA}$ (\ref{emntbatri},\ref{emntbanli}) [black dotted line] all taken from the range $(-t/2,t/2)$. The asymptotic form $g_{{\rm nb},0}$ (\ref{gnbzero}) is also shown [red dashed line]. Inset shows the deviation of the actual degeneracy from $g_{{\rm nb},0}$ in a~magnified vertical scale.
}
\end{figure}

To further illustrate a possible role of number-theoretic degeneracies in graphene nanoflakes we consider the energy levels recently found by Rozhkov and Nori for a tight-binding Hamiltonian of the equilateral triangle with armchair edges  \cite{Roz10}
\begin{multline}
\label{emntbatri}
  E_{m,n,\pm}^{\rm TBA}= \pm{}t\left\{ 
    3 + 2\cos\left[\frac{2\pi{}\tilde{n}}{3(N_a\!+\!1)}\right]  
  \right. \\
  \left. 
    + 2\cos\left[\frac{2\pi{}\tilde{m}}{3(N_a\!+\!1)}\right]
    + 2\cos\left[\frac{2\pi{}(\tilde{n}+\tilde{m})}{3(N_a\!+\!1)}\right]
  \right\}^{1/2},
\end{multline}
where  $N_a\equiv{}3H/(2a)$ (such that $N_{\rm tot}=3N_a(N_a+1)$, see Ref.\ \cite{sizefoo}), $\tilde{m}\equiv{}N_a-m$, and $\tilde{n}\equiv{}N_a+n$. For $1\leqslant{}m\leqslant{}n\ll{}N_a$ one gets $3(N_a+1)\simeq{}\sqrt{3N_{\rm tot}}$ and $E_{m,n,\pm}^{\rm TBA}\simeq{}E_{m,n,\pm}^{\rm Dirac}$, restoring the energy levels of a Dirac cavity (\ref{emndiractri}). Furthermore, expanding Eq.\ (\ref{emntbatri}) in series and keeping the terms up to the order of $\sim{}m^rn^{(3-r)}/(N_a)^3$, with the integer $0\leqslant{}r\leqslant{}3$, we can write
\begin{equation}
  \label{emntbanli}
  \left(\frac{E_{m,n,\pm}^{\rm TBA}}{\Delta}\right)^2\simeq
  \left(\epsilon_{k,l}\right)^2
  - \frac{\pi}{\sqrt{3}}\,\frac{l^2(l-k)}{N_a+1},
\end{equation}
where the notation of Eq.\ (\ref{emndiracdim}) is used. The last term in Eq.\ (\ref{emntbanli}) represents so-called {\em trigonal warping} of the dispersion relation appearing in the vicinity of each Dirac points in graphene \cite{And07}. It is clear, that the quantity $(E_{m,n,\pm}^{\rm TBA})^2-(E_{m',n',\pm}^{\rm TBA})^2$ is not an integer multiplicity of $\Delta^2$ for arbitrary $(m',n')\neq(m,n)$, and thus the quantization rule for $S$ (\ref{squant}) no longer applies. 

Additionally, the trigonal warping appears to be the reason for which number-theoretic degeneracies are totally absent in spectra of tight-binding Hamiltonians for graphene nanoflakes. In Fig.\ \ref{ndegatri}, we plot the average degeneracy ($g_{\rm nb}$) of energy levels from the interval $(-t/2,t/2)$ obtained from Eqs.\ (\ref{emndiractri}) [blue solid line], and (\ref{emntbatri},\ref{emntbanli}) [black dotted line] as a function of $N_{\rm tot}\leqslant{}10^8$. (Notice that as we focus on number-theoretic degeneracies, the twofold degeneracy $E_{m,n,\pm}^{\rm TBA}=E_{n-m,n,\pm}^{\rm TBA}$ of each energy level for which $m\neq{}n$ is not taken into account.) The asymptotic form of $g_{\rm nb}$ for large $N_{\rm tot}$ in the presence of number-theoretic degeneracies \cite{Ber81}
\begin{equation}
  \label{gnbzero}
  g_{{\rm nb},0}\propto\sqrt{\log\left(N_{\rm tot}/{\rm const}\right)}
\end{equation}
is divergent for $N_{\rm tot}\rightarrow\infty$ and depicted in Fig.\ \ref{ndegatri} [red dashed line]. For Dirac cavities, energies $E_{m,n,\pm}^{\rm Dirac}$ (\ref{emndiractri}) show the first number-theoretic degeneracy at $N_{\rm tot}=2610$, for which $g_{\rm nb}=\frac{17}{16}>1$. For larger $N_{\rm tot}$, $g_{\rm nb}\simeq{}g_{{\rm nb},0}$ with the accuracy better than $1\%$ if $N_{\rm tot}\gtrsim{}10^6$. In contrast, neither exact energies of tight-binding Hamiltonian for triangular nanoflakes $E_{m,n,\pm}^{\rm TBA}$ (\ref{emntbatri}) nor these given by the approximating Eq.\ (\ref{emntbanli}) show number-theoretic degeneracies (i.e., $g_{\rm nb}=1$ for any $N_{\rm tot}$).

\section{Transition GUE-GOE for $4\times{4}$ real symmetric matrices %
  \label{tra4mat}}

In this Appendix, we analyze numerically a simple additive model of random matrices capable of describing the transition GUE-GOE associated with the splitting of twofold valey degeneracy in graphene. A single-parameter formula approximating the nearest-neighbor spacing distributions $P^{(1)}(S)$ is proposed.

The analysis starts from the model $2N\times 2N$ real symmetric matrix $H(\lambda)$ of the form (\ref{admatm}), where $H^0$ has a block-structure
\begin{equation}
  H^0=\left(\begin{array}{cc}
    A  & B  \\
    -B & A  \\
  \end{array}\right),
\end{equation}
with $A=A^T$ and $B=-B^T$. The elements of each block are independently generated according to a Gaussian distribution with zero mean and the variances $\mbox{Var}(A_{ij})=(1+\delta_{ij})/2N$ and $\mbox{Var}(B_{ij})=(1-\delta_{ij})/2N$. Therefore, $H^0$ can be unitary mapped onto complex Hermitian matrix 
\begin{equation}
  \tilde{H}^0=\left(\begin{array}{cc}
    A+iB  & 0  \\
    0 & A+iB  \\
  \end{array}\right), 
\end{equation}
which has a twofold eigenvalue degeneracy and belongs to GUE. The matrix $V$ in Eq.\ (\ref{admatm}) is now chosen as a $2N\times{}2N$ member of GOE. 

Varying the scaling parameter $\lambda$, we transform the symmetry class of random matrix $H(\lambda)$ from the unitary ($\lambda=0$) to the orthogonal ($\lambda=\infty$), simultaneously splitting the twofold eigenvalue degeneracy. The nearest-neighbor spacing distributions are approximated by
\begin{equation}
  \label{psakap}
  P_{\alpha,\kappa}^{(1)}(S) = \frac{P_{\rm odd}(\alpha;S) 
    + P_{\rm even}(\beta,\kappa;S)}{2}
\end{equation}
with $P_{\rm odd}(\alpha;S)$ and $P_{\rm even}(\beta,\kappa;S)$ given by Eqs.\ (\ref{psodd}) and (\ref{pseven}), respectively. Eq.\ (\ref{psakap}) represents an observation, that the spacings distribution for the random matrix $H(\lambda)$ consists of two contributions (both of the equal weights for large $N$): first from \emph{odd} spacings, separating the levels that are degenerate for $\lambda=0$, and the second from \emph{even} spacings. We further suppose the distribution of odd spacings is well approximate by the Wigner surmise for GOE (\ref{pgoe}), whereas even spacings undergoes the transition GUE-GOE according to the Berry-Robnik formula (\ref{psgoegue}). The relation between parameters $1\leqslant\alpha<\infty$ and $0\leqslant\kappa<\infty$ of the spacing distribution $P_{\alpha,\kappa}(S)$ and the scaling parameter $\lambda$ is to be determined.

\begin{figure}
\centerline{\includegraphics[width=0.9\linewidth]{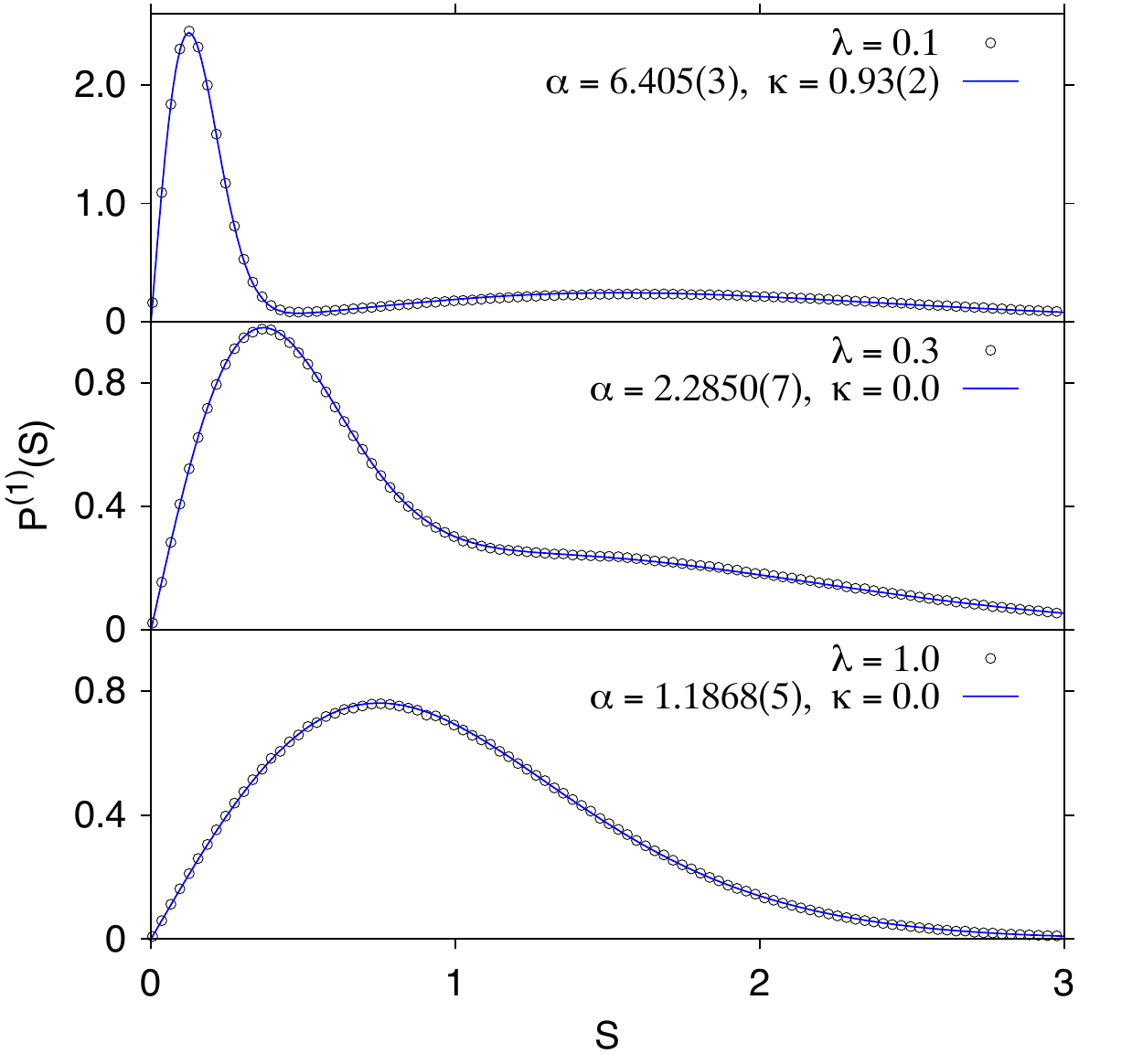}}
\caption{\label{psfit4m}
   Level-spacing distributions averaged over $10^7$ randomly-generated 
   matrices $H(\lambda)$ [datapoints]. 
   The scaling parameter $\lambda$ is varied between the panels. 
   The least-square fitted functions $P_{\alpha,\kappa}^{(1)}(S)$ (\ref{psakap}) 
   are also shown [solid lines].}
\end{figure}

\begin{figure}
\centerline{\includegraphics[width=0.9\linewidth]{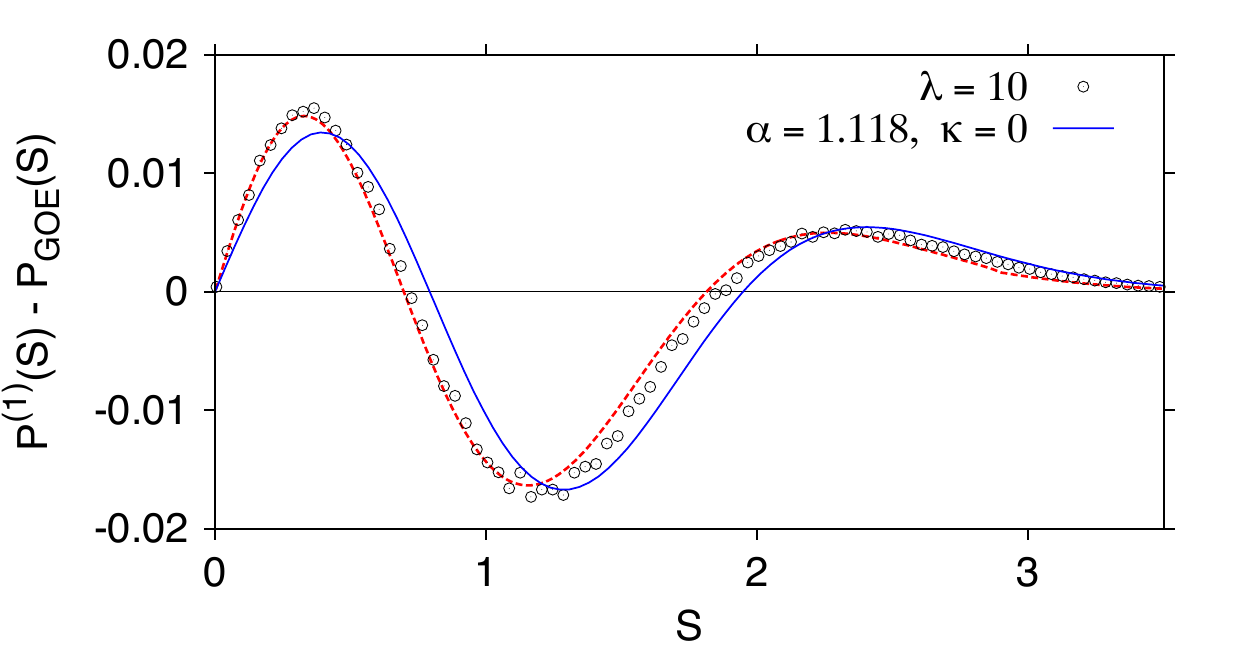}}
\caption{\label{pslam10dev}
  Deviation of the spacing distribution for $10^8$ matrices $H(\lambda)$
  with $\lambda=10$ from the Wigner surmise for GOE (\ref{pgoe}). 
  Blue solid line shows $P_{\alpha,\kappa}^{(1)}(S)$ (\ref{psakap}) fitted to the 
  actual data [points]. Red dashed line corresponds to the asymptotic 
  ($N\rightarrow\infty$) distribution for GOE \cite{Die90}.}
\end{figure}

\begin{figure}
\centerline{\includegraphics[width=\linewidth]{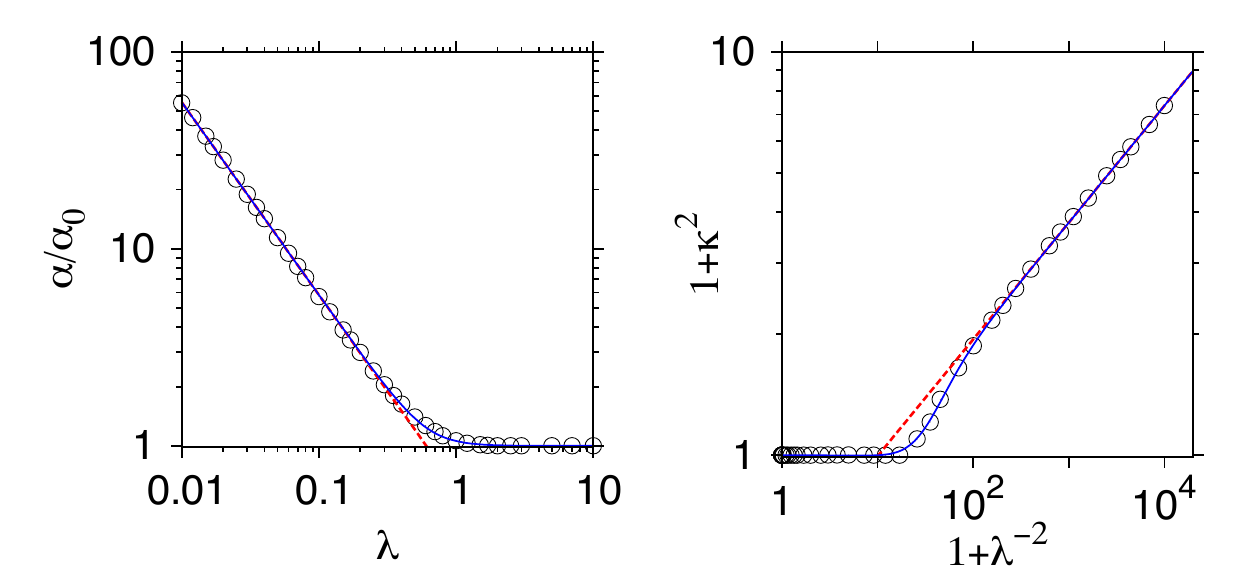}}
\caption{\label{fitparn4}
  Best-fitted parameters of $P_{\alpha,\kappa}^{(1)}(S)$ (\ref{psakap}) [datapoints]
  and the empirical functions  $\alpha=\overline\alpha(\lambda)$ 
  (\ref{alpemp}) and $\kappa=\overline\kappa(\lambda)$ (\ref{kapemp}) 
  [solid lines]. 
  Dashed lines depict the asymptotic forms (\ref{asympara}).}
\end{figure}

We test numerically the formula (\ref{psakap}) for $N=2$ and $\lambda=10^{-2}-10^2$. For each value of $\lambda$, an ensemble of $10^7-10^8$ pseudorandom matrices was generated, each $4\times 4$ matrix was diagonalized, and a histogram of spacings distribution $P^{(1)}(S)$ was obtained. To truncate the large-$N$ limit, in which the contributions from odd and even spacings have equal weights, we took each first and third spacing with the weight $1/4$, whereas second spacing weight was set to $1/2$. Subsequently, the function $P_{\alpha,\kappa}(S)$ was fitted to the numerical data within the least-squares method. Selected examples are presented in Fig.\ \ref{psfit4m}. Remarkably, the parameter $\alpha$ does not approach $1$ for large $\lambda$. This is because Wigner surmise $P_{\rm GOE}(S)$ (\ref{pgoe}) represents the spacing distribution exactly for $2\times 2$ GOE matrices only. For $4\times 4$ matrices, formula (\ref{psakap}) with $\alpha=\alpha_0\simeq{}1.118$ and $\kappa=0$ appears to provide much better approximation of actual spacings distribution $P(S)$ than the Wigner surmise (see Fig.\ \ref{pslam10dev}). 

The collection of best-fitted distributions $P_{\alpha,\kappa}^{(1)}(S)$ allows us to propose empirical functions $\alpha=\overline\alpha(\lambda)$ and $\kappa=\overline\kappa(\lambda)$. The first function reads
\begin{equation}
\label{alpemp}
  \overline\alpha(\lambda) = \alpha_0\left(\frac{\lambda_1}{\lambda}\right)^b
  \left[\,1+\left(\frac{\lambda}{\lambda_1}\right)^c\,\right]^{b/c},
\end{equation}
with
\begin{gather}
  \alpha_0=1.118(1),\ \ \ \lambda_1=0.607(3),\ \ \ b=0.978(2), \nonumber \\ 
  c=3.2(4),
\end{gather}
where standard deviations obtained from least-square fitting are specified in the parenthesis. (It is worth to stress, that the value of a~parameter $\alpha_0$ was found directly from spacing distributions corresponding to $\lambda\gtrsim{}10$ for which $\kappa\simeq{}0$.) The second function is given by
\begin{align}
\label{kapemp}
  \overline\kappa(\lambda) &= \sqrt{ \left(
      \frac{1+\lambda^{-2}}{1+\lambda_c^{-2}}
    \right)^{\overline\gamma(\lambda)} - \,1\, }, \nonumber \\
  \overline\gamma(\lambda) &= \gamma_0 \left[
    1 + \left(\frac{1+\lambda_2^{-2}}{1+\lambda^{-2}}\right)^\delta
  \right]^{-1},
\end{align} 
with
\begin{gather}
  \gamma_0=0.2898(5),\ \ \ \lambda_c=0.329(2),\ \ \ \delta=2.6(1), \nonumber \\
  \lambda_2=0.178(2).
\end{gather}
The functions  $\overline\alpha(\lambda)$ and $\overline\kappa=\kappa(\lambda)$ are plot in Fig.\ \ref{fitparn4} (solid lines) together with actual datapoints used for the fitting. The asymptotic expressions for small $\lambda$
\begin{equation}
\label{asympara}
  \overline\alpha \simeq \alpha_0(\lambda_1/\lambda)^b, \ \ \ \ \ \ 
  \overline\kappa \simeq 
    \left(\frac{1+\lambda^{-2}}{1+\lambda_c^{-2}}\right)^{\gamma_0}
\end{equation}
are also shown in Fig.\ \ref{fitparn4} (dashed lines). Substituting  $\alpha=\overline\alpha(\lambda)$ and $\kappa=\overline\kappa(\lambda)$ to the formula (\ref{psakap}) we obtain a single-parameter function $P_{\overline\alpha,\overline\kappa}^{(1)}(\lambda;S)$ given by Eq.\ (\ref{ps1barak}) in the main text. Thus, the construction of an empirical formula for the nearest-neighbor spacing distribution of $H(\lambda)$ is complete.



\begin{thebibliography}{}
\bibitem{zeeman}
Hereinafter, we neglect the Zeeman term in the Hamiltonian. Such a term leads to the energy level spliting of $\Delta{E}_{\rm Z}=g\mu_B{B}$ (with $g\simeq{2}$ and $\mu_B=e\hbar/(2m_e)$ the Bohr magneton). On the other hand, typical level separation in graphene nanoflake of $\mathcal{A}\simeq(40\,\mbox{nm})^2$ area, for the energies not exceeding $E_{\rm max}=1\,$eV, obeys $\Delta{E}_{\cal A}\gtrsim\pi(\hbar{}v_F)^2/({\cal A}E_{\rm max})\simeq{10}^{-3}\,$eV. At the field of $B=1\,$T we get $\Delta{E}_{\rm Z}/\Delta{E}_{\cal A}\lesssim{0.1}$, so the Zeeman spliting shall not visibly affect spectral statistics discussed in the paper.
\bibitem{sizefoo}
The total number of lattice sites $N_{\rm tot}$ is related to the system height $H$ via $N_{\rm tot}=\frac{4}{3}(H/a)^2+\alpha{}(H/a)+\beta$ for triangles or $N_{\rm tot}=2(H/a)^2+\gamma$ for hexagons. The coefficients $(\alpha,\beta,\gamma)\equiv(2,0,3/2)$ for armchair edges, $(4/\sqrt{3},-2,0)$ for zigzag edges, or $(2\sqrt{3},-6,-3/2)$ for Klein edges. For triangles, $H/a$ is integer for armchair edges, or is given by an integer multiplicity of $\sqrt{3}/2$ for zigzag or Klein edges. For hexagons with armchair, zigzag, or Klein edges, $H/a$ is half-odd integer, integer multiplicity of $\sqrt{3}$, or half-odd integer multiplicity of $\sqrt{3}$ (respectively).
\bibitem{pksdef}
Notice that the definition of $P^{(k)}(S)$ implies $\int_0^{\infty}SP^{(k)}(S)dS=\int_0^{\infty}P^{(k)}(S)dS=1$ for $k=1,2,\dots$.
\bibitem{alambdapfq}
Using the exponential integral $\mbox{Ei}(x)$ (see Ref.\ \cite{Abram}, Chapter 5), the complementary error function $\mbox{erfc}(x)\equiv{}1-\mbox{erf}(x)$ and the generalized hypergeometric function $_pF_q(a_1,\dots,a_p;b_1,\dots,b_q;z)$ \cite{Olv10} one can also write Eq.\ (\ref{alambda}) in the form 
$
a(\lambda)= \sqrt{\frac{2}{\pi}}\lambda+\exp\left(\frac{\lambda^2}{2}\right)
\mbox{erfc}\left(\frac{\lambda}{\sqrt{2}}\right)
-\frac{\lambda^2}{2}\mbox{Ei}\left(\frac{\lambda^2}{2}\right)
+\sqrt{\frac{2}{\pi}}\lambda^3\,{}_2F_2\left(\frac{1}{2},1;
  \frac{3}{2},\frac{3}{2};\frac{\lambda^2}{2}\right).
$
\bibitem{psazfoo}
We have varied the impurity concentration in a range $5.4\times{}10^{-4}\leqslant{}N_{\rm imp}/N_{\rm tot}\leqslant{}0.011$ leaving the other simulation details same as in Ref.\ \cite{csklfoo}.
\bibitem{lbfitfoo}
We have varied the impurity concentration in a range $2.1\times{}10^{-4}\leqslant{}N_{\rm imp}/N_{\rm tot}\leqslant{}0.045$ taking the other parameters same as in Ref.\ \cite{csklfoo}. The distributions $P^{(1,2)}(S)$ used to determine $\lambda_{\rm fit}$ were obtained by averaging over $300-400$ independent disorder realizations for small systems ($N_{\rm tot}\lesssim{}10^4$) or by taking just one disorder realization for large systems ($N_{\rm tot}\gg{}10^4$).
\bibitem{del3foo}
The asymptotic forms for $L\gg{}1$ are $\Delta_3(L)\simeq{}\pi^{-2}[\ln(2\pi{}L)+\gamma-\frac{5}{4}-\pi^2/8]$ for GOE and $\Delta_3(L)\simeq{}(2\pi^2)^{-1}[\ln(2\pi{}L)+\gamma-\frac{5}{4}]$ for GUE, with the Euler constant $\gamma\simeq{}0.577$ \cite{MehA40}. The above forms reproduce the theoretical values of $\Delta_3(L)$ given Eqs.\ (\ref{del3th},\ref{ysgue}) up to the second decimal place if $L\gtrsim{}10$.
\bibitem{psnvfoo}
We took $N_{\rm imp}/N_{\rm tot}=0.034$, with $N_{\rm tot}+N_{\rm vac}=32758$ for ${\cal A}\simeq{}(120\,a)^2$ [or $N_{\rm tot}+N_{\rm vac}=8278$ for ${\cal A}\simeq{}(60\,a)^2$], and the other parameters same as in Ref.\ \cite{csklfoo}.
\bibitem{lamzefoo}
The relative valley splitting following from the level-spacing distribution $P_{\overline{\alpha},\overline{\kappa}}^{(1)}(\lambda_{\rm fit};S)$ (\ref{ps1barak}) can be estimated as $\langle\rho(E)\rangle\Delta{}E_{K,K'}\sim{}1/\overline{\alpha}\simeq{}1.46\times\lambda_{\rm fit}^{0.98}$, where we have used the asymtotic form of $\overline{\alpha}$ (\ref{asympara}) for $\lambda=\lambda_{\rm fit}\ll{}1$.
\bibitem{laminfoo}
The Wigner surmise $P_{\rm GOE}(S)$ (\ref{pgoe}) represents the exact expression for spacing distributions only for $2\times{}2$ members of GOE, but it approximates $P_{\overline\alpha,\overline\kappa}^{(1)}(\infty;S)$ within the accuracy close to $1\%$.
\bibitem{eguefoo}
We have used the Wigner surmise for GUE (\ref{pgue}). More generally, $E(S)=\int_0^{\infty}du\int_0^{\infty}dv\,P(S+u+v)$.
\bibitem{kldefoo}
Notice that the definitions of $k$ and $l$ imply (for armchair edges) $0\leqslant{}k\leqslant{}l-2$ and $k\,\mbox{mod}\,2=l\,\mbox{mod}\,2$ (i.e., integers $k$ and $l$ are both even or both odd).
\end{thebibliography}
\end{document}